\documentclass[a4paper,prd,twocolumn,aps,preprintnumbers,showpacs]{revtex4}
\usepackage{amssymb}
\usepackage{graphicx}
\usepackage{dcolumn}
\usepackage{bm}
\usepackage[colorlinks=true,
            linkcolor=blue,
            citecolor=blue]{hyperref}
\usepackage{slashed}
\usepackage{enumitem}
 
\def \theequation {\arabic{equation}}             
\def \be  {\begin{equation}}
\def \ee  {\end{equation}}
\def \beq  {\begin{equation}}
\def \eeq {\end{equation}}
\def \ba  {\begin{eqnarray}}
\def \ea  {\end{eqnarray}}
\def \baa {\begin{eqnarray*}}
\def \eaa {\end{eqnarray*}}
\def \bb  {\begin{thebibliography}}
\def \eb  {\end{thebibliography}}

\def \nn {\nonumber}

\def \lab #1 {\label{#1}}

\def \CO {{\cal O}}

\def \fracs #1#2 {\mbox{\small $\frac{#1}{#2}$}}

\def \bin #1#2 {{\left({#1}\atop{#2}\right)}}
\def\lapproxeq{{\ \lower 0.6ex \hbox{$\buildrel<\over\sim$}\ }}
\def\gapproxeq{{\ \lower 0.6ex \hbox{$\buildrel>\over\sim$}\ }}

\def\hepph  #1 {{hep-ph/#1 }}

\begin{document}

\title{Single-Spin Asymmetries in $W$ Boson Production at Next-to-Leading Order}
\author{Felix Ringer, Werner Vogelsang}
\affiliation{Institute for Theoretical Physics, 
T\"ubingen University, Auf der Morgenstelle 14, 72076 T\"ubingen, Germany}
\begin{abstract}
We present an analytic next-to-leading order QCD calculation of the partonic cross sections 
for single-inclusive lepton production in hadronic collisions, when the lepton originates from
the decay of an intermediate electroweak boson and is produced at high transverse momentum. 
In particular, we consider the case of incoming longitudinally polarized protons for which 
parity-violating single-spin asymmetries arise that are exploited in the $W$ boson program 
at RHIC to constrain the proton's helicity parton distributions. Our calculation enables a very
fast and efficient numerical computation of the relevant spin asymmetries at RHIC, which is
an important ingredient for the inclusion of RHIC data in a global analysis of nucleon helicity 
structure. We confirm the validity of our calculation by comparing with an existing code that treats
the next-to-leading order cross sections entirely numerically by Monte-Carlo integration techniques. 
We also provide new comparisons of the present RHIC data with results for some of the sets
of polarized parton distributions available in the literature.
\end{abstract}

\date{\today}
\pacs{12.38.Bx, 14.70.-e, 13.88.+e}
\maketitle

\section{Introduction \label{sec:intro}}

The $W$ physics program at RHIC~\cite{Aschenauer:2015eha} is dedicated to providing new
insights into the helicity structure of the proton. It exploits the violation of parity in the weak
interactions, which gives rise to {\it single}-longitudinal spin asymmetries in proton-proton collisions.
The main focus is on the production of $W$ bosons, identified by their subsequent decay into a lepton 
pair. The charged lepton (or antilepton) is observed. From the corresponding cross sections for the
various helicity settings $(++),(+-),(-+),(--)$ of the two incoming protons one defines the spin asymmetry
\be\label{eq:asymW}
A_L^{W^\pm}\equiv \frac{d\sigma^{++}+d\sigma^{+-}-(d\sigma^{-+}+d\sigma^{--})}{
d\sigma^{++}+d\sigma^{+-}+d\sigma^{-+}+d\sigma^{--}}\equiv\frac{d\Delta\sigma}{d\sigma}\;.
\ee
As one can see, one takes the difference of cross sections for positive and negative
helicities of one proton, while summing over the polarizations of the other. 
The STAR collaboration at RHIC has published rather extensive and precise 
data on $A_L^{W^\pm}$ last year~\cite{Adamczyk:2014xyw}, and new precise
mid-rapidity data from PHENIX have just appeared~\cite{phenix}. Earlier measurements 
were reported by both PHENIX~\cite{Adare:2010xa} and STAR~\cite{Aggarwal:2010vc}. 
Data sets with even higher statistics and kinematic coverage are expected in the near future. 
Typically, the data are presented at fixed rapidity of the charged lepton, which by convention is
counted as positive in the forward direction of the polarized proton. 

It has long been recognized~\cite{Bourrely:1993dd,Bunce:2000uv} that $A_L^{W^\pm}$ offers
excellent sensitivity to the individual helicity parton distributions $\Delta u,\,\Delta\bar{u},\,\Delta d,\,\Delta\bar{d}$
of the nucleon, where
\be\label{eq:polPDF}
\Delta f(x,Q^2)\equiv f^+(x,Q^2)-f^-(x,Q^2)\;,
\ee
with $f^+$ ($f^-$) denoting the distribution of parton $f$ with positive (negative) helicity in a parent proton
with positive helicity. The distributions are functions of the longitudinal momentum fraction $x$ of the parton
and of a ``resolution'' scale $Q$. Information on $\Delta u,\,\Delta\bar{u},\,\Delta d,\,\Delta\bar{d}$ is also 
accessible in (semi-inclusive) deep-inelastic lepton scattering 
(DIS)~\cite{Airapetian:2004zf,Alekseev:2010hc,deFlorian:2014yva,deFlorianpPDF,Leader:2010rb}. 
The key advantages of $W$ boson production are that (i) it is characterized by momentum scales of the 
order of the $W$ mass which are much higher than those presently relevant in DIS and hence deeper
in the perturbative domain, (ii) it does not rely on the knowledge of hadronic fragmentation functions, thanks
to its clean leptonic final state. In any case, information from the $W$ program at RHIC is complementary 
to that from DIS. 

The main concept behind the RHIC measurements can be easily summarized:
For $W^-$ production, taking into consideration only the dominant $\bar{u}d \to W^-$ 
subprocess, the spin-dependent cross section in the numerator of the asymmetry
in Eq.~(\ref{eq:asymW}) is found to be proportional to the combination
\be
\Delta \bar{u}(x_1)d(x_2)(1-\cos\theta)^2-
\Delta d(x_1) \bar{u}(x_2)(1+\cos\theta)^2\, ,
\label{eq:w-lo}
\ee
where for simplicity we have not written out the straightforward convolutions over the 
parton momentum fractions. $\theta$ is the polar angle of the negatively charged decay 
lepton in the partonic center-of-mass system, with $\theta > 0$ in the forward direction 
of the polarized parton. In the backward region of the lepton,
one has $x_2\gg x_1$ and $\theta\gg\pi/2$, so that the
first term in Eq.~(\ref{eq:w-lo}) strongly dominates. Since the denominator
of $A_L$ is proportional to $\bar{u}(x_1)d(x_2)(1-\cos\theta)^2+
d(x_1) \bar{u}(x_2)(1+\cos\theta)^2$, the asymmetry then provides a
clean probe of $\Delta\bar{u}(x_1)/\bar{u}(x_1)$ at medium
values of $x_1$. By similar reasoning, in the forward lepton region the 
second term in Eq.~(\ref{eq:w-lo}) dominates, giving access
to $-\Delta d(x_1)/d(x_1)$ at relatively high $x_1$. 

For $W^+$ production, within the same approximation, the 
spin-dependent cross section is proportional to
\be
\Delta \bar{d}(x_1)u(x_2)(1+\cos\theta)^2-
\Delta u(x_1) \bar{d}(x_2)(1-\cos\theta)^2\, .
\label{eq:w+lo}
\ee
Here the distinction of the two contributions by considering
backward or forward lepton scattering angles is less clear-cut than
in the case of $W^-$ because of the reversal of the factors
$(1\pm \cos\theta)^2$ relative to~(\ref{eq:w-lo}), which always suppresses
the dominant combination of parton distributions. Therefore,
both terms in (\ref{eq:w+lo}) will compete. Nonetheless, the $W^+$ 
measurements at RHIC are of course of great value in the context
of a global analysis of the helicity distributions.

Given the importance of $A_L^{W^\pm}$ for constraining nucleon helicity
structure, there has been a lot of activity on the calculation of higher-order
QCD corrections to the relevant spin-dependent cross sections. Closed analytic
expressions for next-to-leading order (NLO) corrections to polarized $W$ boson
production were derived in Refs.~\cite{Kamal:1997fg,Gehrmann:1997ez}, 
with extensions to all-order resummations in~\cite{Weber:1993xm,Mukherjee:2006uu}.
In these papers, direct observation of the $W$ boson and its kinematics was
assumed, which simplifies the calculation considerably but is 
not really applicable to the measurements at RHIC. The proper
lepton decay kinematics was taken into account in three further 
studies~\cite{Nadolsky:2003ga,deFlorian:2010aa,vonArx:2011fz}. 
The first two of these include the contributions by intermediate $Z$ bosons and photons as well, 
which may also give rise to charged leptons and provide a background to the lepton signal from
$W$ boson decay when the detectors are not hermetic. Reference~\cite{Nadolsky:2003ga} 
additionally derives and implements the resummation of large logarithms in the transverse 
momentum of the intermediate $W$ boson. 

In the calculations~\cite{Nadolsky:2003ga,deFlorian:2010aa,vonArx:2011fz} 
the NLO corrections were obtained numerically in the context of a Monte-Carlo
integration routine. The resulting computer codes are very flexible in the sense that
kinematic cuts on lepton or recoil jet variables can be easily implemented. Those from 
Refs.~\cite{Nadolsky:2003ga} and~\cite{deFlorian:2010aa} are known 
as RHICBOS and CHE, respectively, and have found wide use in comparisons 
to RHIC data. On the other hand, the Monte-Carlo integration based codes are rather 
demanding in terms of CPU time. This becomes a significant drawback when one 
wants to perform a global analysis of the helicity distributions from the RHIC
data~\cite{deFlorian:2014yva,deFlorianpPDF,Nocera:2014gqa}. 
Such an analysis typically requires many thousands of computations of the
spin asymmetry. Clearly, a fast and stable evaluation at NLO is highly desirable
in this context. 

In this paper, we derive analytic expressions for the NLO spin-dependent partonic cross 
sections for electroweak boson production, {\it including} their leptonic decay. More 
precisely, we consider the cross sections directly as single-inclusive lepton ones, 
$\vec{p}p \rightarrow \ell^{\pm}X$, where transverse momentum and rapidity of the 
charged lepton are observed, precisely as is the case at RHIC. We note that a corresponding
calculation in the unpolarized case has been presented a long time ago~\cite{Aurenche:1980tp}.
We present a new program that produces NLO results for the single-spin asymmetries relevant
at RHIC and outruns the Monte-Carlo based codes by about two orders of magnitude in CPU time. 
We also include the background reactions involving $Z$ bosons and photons. We expect
our program to become a useful tool for global analyses of RHIC data based on 
Mellin-moment~\cite{deFlorian:2014yva,deFlorianpPDF,Stratmann:2001pb} or neural-network~\cite{Nocera:2014gqa} 
techniques. We also use our new code to present comparisons of the present RHIC data
to NLO predictions for a variety of sets of helicity parton distributions.

In Sec.~\ref{sec:WpNLO} we discuss the technical details of our NLO calculation. Section~\ref{sec:Wpnumericalresults}
presents our phenomenological results, where we also perform comparisons with the CHE code 
of~\cite{deFlorian:2010aa}. Finally, we conclude in Section~\ref{sec:Wpconclusions}.

\section{Next-to-leading order calculation \label{sec:WpNLO}}

\subsection{Framework and outline of the NLO calculation}

We consider the single-inclusive process $\vec{p} p\rightarrow \ell+X$, where $\ell$ denotes the 
charged lepton (or antilepton) resulting from production and decay of a $W$ boson. As discussed in
the Introduction, charged leptons can of course also be produced by an intermediate photon or $Z$ boson 
which, subject to the experimental selection criteria, gives rise to a background. We hence perform all
our calculations also for $\gamma$ and $Z$ production and $\gamma Z$ interference. For the sake
of simplicity we will, however, present details of our calculation and explicit results only for the most
interesting $W$ boson case, and just highlight a few features specifically relevant for intermediate 
$\gamma$ and $Z$.

We denote the momenta of the incoming protons and the produced charged lepton by $P_A,P_B,p_\ell$, respectively. 
Using factorization~\cite{Collins:1989gx}, we write the polarized hadronic cross section $d\Delta\sigma$ 
which appears in the numerator of Eq.~(\ref{eq:asymW}) in terms of convolution integrals of polarized and 
unpolarized parton distributions $\Delta f_a(x_a,\mu_F)$, $f_b(x_b,\mu_F)$ 
and the perturbative hard-scattering partonic cross sections $d\Delta\hat{\sigma}_{ab}$: 
\ba\label{eq:factorization}
d\Delta\sigma &=& \sum_{a,b}\int dx_a dx_b \, \Delta f_a(x_a,\mu_F)\, f_b(x_b,\mu_F)\nn\\[2mm]
&&\times\, d\Delta \hat{\sigma}_{ab}(x_a P_A,x_bP_B,p_\ell,\mu_R,\mu_F)\;,
\ea
where
\be\label{eq:asymWpart}
d\Delta \hat{\sigma}_{ab}\,\equiv\,\frac{1}{4}\left[d\hat{\sigma}^{++}+d\hat{\sigma}^{+-}-(d\hat{\sigma}^{-+}+
d\hat{\sigma}^{--})\right]\;.
\ee
The superscripts on the right refer to parton helicities, so that the helicities of the 
second parton $b$ are summed over, while he helicity difference is taken for parton $a$.
The sum in Eq.~(\ref{eq:factorization}) runs over quarks, antiquarks and the gluon, and the parton distributions 
are evaluated at the factorization scale $\mu_F$. The partonic cross sections also depend
on a renormalization scale $\mu_R$. The fractions of the parent hadrons' momenta
carried by the scattering partons are denoted by $x_a$ and $x_b$. An analogous expression for the 
unpolarized cross section $d\sigma$ appearing in the denominator of Eq.~(\ref{eq:asymW}) is obtained 
by using only unpolarized parton distributions and the corresponding unpolarized partonic cross sections, 
defined by averaging over the helicities of both incoming partons. 

Due to the pure $V-A$ structure of the $Wq\bar{q}'$ vertex, and because of conservation of quark helicity
at the vertex, the spin-dependent partonic cross section for an incoming polarized quark is
just the negative of the corresponding unpolarized cross section, while for an incoming polarized 
{\it anti}-quark it is the same as the unpolarized one:
\ba\label{crsecrel}
d\Delta \hat{\sigma}_{qb}&=&-d \hat{\sigma}_{qb}\hspace*{1cm}(b=\bar{q}',g)\,,\nn\\[2mm]
d\Delta \hat{\sigma}_{\bar{q}b}&=&d \hat{\sigma}_{\bar{q}b}\hspace*{1.28cm}(b=q',g)\,.
\ea
Note that no such relation occurs for incoming polarized gluons. In case of $\gamma$ and/or $Z$
exchange, relations~(\ref{crsecrel}) do not hold.

We now introduce the variables
\be\label{STU}
S\equiv (P_{A}+P_{B})^2\,,\;\;T \equiv (P_{A}-p_\ell)^2\,, \;\; U\equiv (P_{B}-p_\ell)^2\;,
\ee
and
\be\label{VW}
V  \equiv  1+\frac{T}{S}, \enspace  W \equiv \frac{-U}{S+T}\;.
\ee
The lepton's transverse momentum $p_T$ and its center-of-mass system 
rapidity $\eta$ are related to these variables by
\be
V=1-\frac{p_T}{\sqrt{S}}{\mathrm{e}}^{-\eta}\,,\;\;VW=\frac{p_T}{\sqrt{S}}{\mathrm{e}}^\eta\;.
\ee
We furthermore introduce the partonic variables corresponding to Eqs.~(\ref{STU}),(\ref{VW}):
\ba\label{eq:vw0}
&&s\equiv (p_{a}+p_{b})^2, \enspace  t \equiv  (p_a-p_\ell)^2,\enspace u\equiv (p_b-p_\ell)^2\,,\nn\\[2mm]
&&v  \equiv  1+\frac{t}{s}, \enspace w \equiv \frac{-u}{s+t},\enspace
\ea
so that from $p_a=x_a P_A$, $p_b=x_b P_B$ we have
\be
x_a =  \frac{VW}{vw}\,,\enspace x_b=\frac{1-V}{1-v}\,.
\ee
Writing out Eq.~(\ref{eq:factorization}) explicitly to $\CO(\alpha_s)$ in the strong coupling constant, we now obtain
\begin{widetext}
\ba\label{eq:hadronicNLO}
\frac{d^2\Delta\sigma}{dp_Td\eta}  &=&\frac{2}{p_T} \sum_{a,b}\int_{VW}^V dv\int_{VW/v}^1
dw\,\, x_a \Delta f_a(x_{a},\mu_{F})\, x_b f_b(x_{b},\mu_{F})   \nn\\[2mm]
&&\times\, \left[\frac{d\Delta\hat{\sigma}_{ab}^{(0)}(s,v)}{dv} \delta(1-w)+\frac{\alpha_{s}(\mu_R)}{2\pi} 
\frac{d\Delta\hat{\sigma}^{(1)}_{ab}(s,v,w,\mu_F,\mu_R)}{dvdw} \right]\;,
\ea
\end{widetext}
where the $d\Delta\hat{\sigma}_{ab}^{(0)}$ represent the leading-order (LO) contributions and the 
$d\Delta\hat{\sigma}_{ab}^{(1)}$ the NLO ones. 

\begin{figure}[b]
\hspace*{-1mm}
\includegraphics[width=0.47\textwidth]{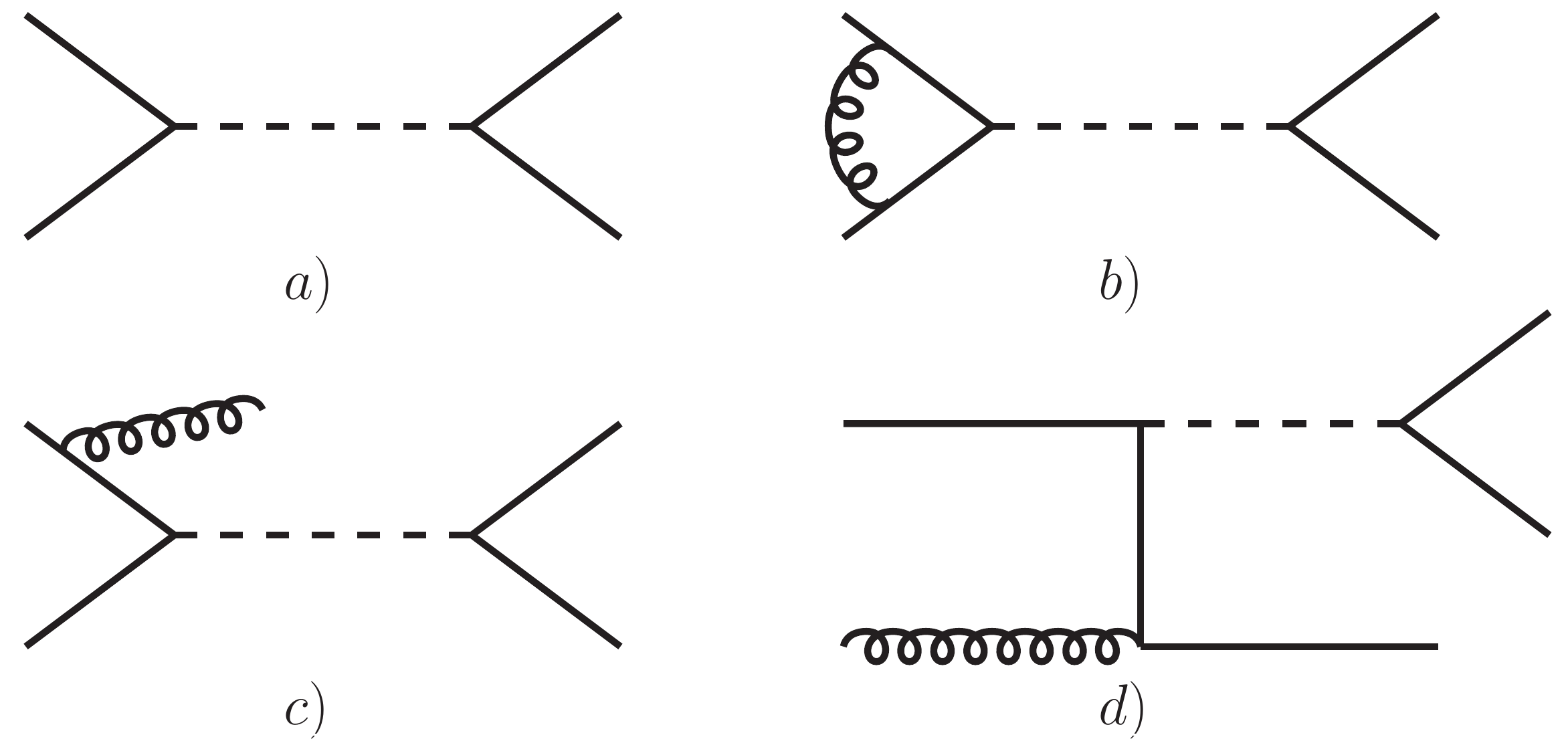}
\caption{\label{fig1_w1}
\sf Feynman diagrams for heavy gauge boson production: a) leading-order, b) NLO virtual correction, 
c) NLO real emission, d) NLO $qg$ scattering. Crossed diagrams are not shown.}
\end{figure}

The only LO partonic process is $q\bar{q}'\rightarrow W\to \ell \nu_\ell$ annihilation, whose Feynman diagram is 
shown in Fig.~\ref{fig1_w1} a). For the NLO correction we have to include the $2\rightarrow 3$ 
real-gluon emission diagrams as well as the virtual corrections to the Born cross section. In addition, 
quark-gluon scattering contributes here as well as a new channel. Some of the relevant NLO Feynman diagrams 
are shown in Fig.~\ref{fig1_w1} (b)-(d). 

For our calculations, we work with a general (axial) vector structure for the 
$Wq\bar{q}'$-vertex of the form 
\be\label{eq:generalvertex}
V_q^\mu = -i\frac{g_W}{2\sqrt{2}} \,U_{qq'}\,\gamma^\mu \,(v_q-a_q\,\gamma_5)\;,
\ee
where $U_{qq'}$ is the appropriate CKM matrix element and $g_W$ the fundamental
weak charge. Likewise, we use a corresponding expression for the 
$W\ell\nu_\ell$-vertex, with vector and 
axial coefficients $v_\ell$ and $a_\ell$ (and, of course, with $U_{qq'}=1$).  
Using such general vertices will help us to keep better track of the couplings in the 
NLO calculation and to obtain an understanding of the underlying structure. 
Also, it allows us to extend our calculation to the case of $\gamma$ or $Z$ boson exchange 
(for $\gamma Z$ interference one needs to introduce an even more general vertex
structure that allows different couplings in the amplitude and its complex conjugate). 
The case of a $W$ boson is recovered by setting $v_q=a_q=1$ and $v_\ell=a_\ell=1$. 

As is very well known, various types of singularities appear at intermediate stages of the NLO 
calculation. To treat these, we choose dimensional regularization with $d=4-2\varepsilon$ dimensions. 
This means that we have to deal with subtleties that occur in Dirac traces involving 
$\gamma_5$ or in the presence of the Levi-Civita tensor $\epsilon^{\mu\nu\rho\sigma}$ 
when $d\neq 4$. $\gamma_5$ appears 
in the $Wq\bar{q}'$-vertex (see~(\ref{eq:generalvertex})) and also acts as projection operator
onto definite helicity states of incoming quarks or antiquarks. Likewise, the Levi-Civita tensor
projects onto gluon helicity states. We adopt the 't Hooft, Veltman, Breitenlohner, Maison (HVBM) 
scheme of~\cite{'tHooft:1972fi,Breitenlohner:1977hr}, which basically recognizes the four-dimensional 
nature of $\gamma_5$ and $\epsilon^{\mu\nu\rho\sigma}$, separating the usual four
space-time dimensions from the additional $d-4=-2\varepsilon$ spatial ones. 
Technically, we compute Dirac traces using the {\sc Tracer} package of~\cite{Jamin:1991dp}. 
We also follow Refs.~\cite{Buras:1989xd,Ciuchini:1993fk} to use a symmetrized version of the
$W$-fermion vertex. 

Because of the distinction between four- and $(d-4)$-dimensional subspaces in the HVBM scheme,
the squared matrix elements for the partonic processes will contain 
regular $d$-dimensional scalar products of the external momenta, but also additionally $(d-4)$-dimensional ones. 
The latter have to be properly taken into account when the phase space integration is performed. 
As it turns out, for the unpolarized cross sections all such additional terms are either absent or 
integrate to zero, {\it i.e.} are of $\CO(\varepsilon)$ after phase space integration. However, in the polarized case, 
they do contribute, and in fact a finite additional subtraction is required in the procedure of factorization of
collinear singularities in order to maintain relations such as~(\ref{crsecrel}) beyond LO. The deeper
reason for this is that the $\gamma_5$ and $\epsilon^{\mu\nu\rho\sigma}$ definitions
of~\cite{'tHooft:1972fi,Breitenlohner:1977hr},
although algebraically consistent, cause violation of helicity conservation at fermion-boson
vertices, which has to be corrected for. Since this is very well established in the literature 
(see, for example, Refs.~\cite{Gordon:1993qc,Vogelsang:1996im,Jager:2002xm}) we shall not 
go into any further detail here but only mention the salient features when they become relevant
in the course of the calculation.

\subsection{Born-level cross section}\label{ssec:Born}

Thanks to~(\ref{crsecrel}), we can easily develop the calculations of the unpolarized and polarized 
cross sections in parallel. Up to the subtleties just mentioned, it is sufficient to present details only
for the unpolarized case. The lowest-order contribution to the cross section comes from the 
$2\rightarrow 2$ scattering process 
$q\bar{q}'\rightarrow\ell\nu_{\ell}$. The diagram is shown in Fig.~\ref{fig1_w1}(a). 
As before, we use  ``$\ell$'' for the observed charged lepton, regardless of its charge. 
We shall see below that it is possible to formulate a partonic cross section in 
this generic way, despite the fact that the ``lepton'' can be either a particle or
an antiparticle. We also always refer to the corresponding neutrino or antineutrino
as the ``neutrino'' and denote it by $\nu_{\ell}$. Since it remains unobserved, we integrate
over its phase space. This leads to an overall factor $\delta(1-w)$ for the Born cross section, so that
\be\label{d1w}
\frac{d^2\hat{\sigma}_{q\bar{q}'}^{(0)}}{dvdw}=\frac{d\hat{\sigma}_{q\bar{q}'}^{(0)}}{dv}\delta(1-w)\;,
\ee
as we have anticipated in~(\ref{eq:hadronicNLO}). 
Using the general vertex structure given in Eq.~(\ref{eq:generalvertex}), we find that 
two combinations of the couplings appear in the expression for the cross section, which are 
given by
\ba\label{C12}
C_1 & = & (v_q^2+a_q^2)(a_\ell^2+v_\ell^2)+4\, a_q a_\ell v_q v_\ell \;, \nn \\[2mm]
C_2 & = &  (v_q^2+a_q^2)(a_\ell^2+v_\ell^2)-4\, a_q a_\ell v_q v_\ell  \; .
\ea
We recall that in case of an exchanged $W^\pm$ boson, we have
$v_q=a_q=v_\ell=a_\ell=1$ and hence always $C_1=8$ and $C_2=0$. 
However, it is useful to keep $C_2$ in the calculation as it allows us to easily
switch between $W^-$ and $W^+$ production. The reason for this becomes 
clear when we write down the unpolarized Born cross section:
\be\label{eq:borncross}
\frac{d\hat{\sigma}_{q\bar{q}'}^{(0)}}{dv}=
\frac{|U_{qq'}|^2s}{8\pi N_c}\left(\frac{G_F M_W^2}{\sqrt{2}} \right)^2
\frac{C_1(1-v)^2+C_2 v^2}{(s-M_W^2)^2+\Gamma_W^2 M_W^2} \;,
\ee
where $N_c=3$, $G_F=\sqrt{2}g_W^2/(8 M_W^2)$ is the Fermi constant, 
and $M_W$ and $\Gamma_W$ are the $W$ boson mass and decay width.
Let us consider now the partonic channel $u\bar{d}\to e^+\nu_e$.
For this indeed Eq.~(\ref{eq:borncross}) provides the correct cross section when
$C_1=8$ and $C_2=0$. In this way the cross section is proportional to $(1-v)^2$,
as required by the $V-A$ structure of the interaction and angular momentum conservation. 
For  $d\bar{u}\to e^-\bar{\nu}_e$, on the other hand, the cross section has to be 
proportional to $v^2$, rather than $(1-v)^2$. This is immediately realized by
interchanging $C_1$ and $C_2$ in Eq.~(\ref{eq:borncross}), and subsequently 
setting again $C_1=8$ and $C_2=0$. Equivalently, and even more simply, we can just 
choose in~(\ref{eq:borncross}) $C_1=8,C_2=0$ for $u\bar{d}\to e^+\nu_e$ and $C_1=0,C_2=8$ 
for $d\bar{u}\to e^-\bar{\nu}_e$ to obtain the correct cross sections. We note that the cross 
sections for the reactions $\bar{d}u\to e^+\nu_e$ and $\bar{u}d\to e^-\bar{\nu}_e$
can be obtained by simple ``crossing'' $t\leftrightarrow u$, or $v\leftrightarrow 1-v$. 
Again this may also be achieved by $C_1\leftrightarrow C_2$. All these considerations
also hold at NLO, where the cross section still depends only on the two combinations 
$C_1$ and $C_2$.

The denominator in Eq.~(\ref{eq:borncross}) represents the standard Breit-Wigner form 
of the propagator. One often also uses the form (see~\cite{Goria:2011wa})
\be
\frac{1}{(s-M_W^2)^2+\Gamma_W^2 s^2/M_W^2}\;,
\ee
which may be obtained from the one given in~(\ref{eq:borncross}) by
the simple rescalings $M_W^2\to M_W^2/(1+\Gamma_W^2/M_W^2)$,
$\Gamma_WM_W\to \Gamma_WM_W/(1+\Gamma_W^2/M_W^2)$ and
multiplication of the cross section by $1/(1+\Gamma_W^2/M_W^2)$.
This also holds at NLO. The numerical difference between these two forms 
of the propagator is very small and negligible for our purposes.

\subsection{Real $2\rightarrow 3$ corrections}

At NLO, we first consider the $2\rightarrow 3$ real-gluon emission process $q\bar{q}'\rightarrow \ell (\nu_{\ell}g)$, 
where the gluon and the neutrino remain unobserved. One of the two relevant Feynman diagrams is shown 
in Fig.~\ref{fig1_w1}(c). All external particles can be considered as massless, so that the kinematics and the 
phase space are as usual for single-inclusive calculations. The three-particle phase space in $4-2\varepsilon$ 
dimensions may be written as~\cite{Gordon:1993qc,Jager:2002xm}
\ba\label{ps3}
\frac{d^2\Phi_{3}}{dvdw} &=& \frac{s}{(4\pi)^{4}\Gamma(1-2\varepsilon)}\left( \frac{4\pi}{s}\right)^{2\varepsilon} \;\nn \\[2mm]
&&  \times \, v^{1-2\varepsilon} \left(1-v\right)^{-\varepsilon}w^{-\varepsilon}\left(1-w\right)^{-\varepsilon} \nn \\[2mm]
 && \times  \int^{\pi}_{0}{d\theta_{1}} \int^{\pi}_{0}d\theta_{2}\sin^{1-2\varepsilon}{\theta_{1}}
 \sin^{-2\varepsilon}{\theta_{2}} \nn\\[2mm]
 && \times \frac{1}{B(1/2,-\varepsilon)}\int_0^1 \frac{dz}{\sqrt{1-z}}z^{-(1+\varepsilon)}\;,
\ea
where $v$ and $w$ have been defined in Eq.~(\ref{eq:vw0}) and where $\theta_1$ and $\theta_2$ are the 
polar and azimuthal angles of the neutrino in the rest frame of the neutrino-gluon pair. The integration 
variable $z$ is specific to the treatment of $\gamma_5$ and $\epsilon^{\mu\nu\rho\sigma}$ in
the HVBM scheme. It is given by  $z\equiv 4\hat{k}^2/(s_{23}\sin^2\theta_1 \sin^2\theta_2)$, where
$s_{23}=sv(1-w)$ and 
$\hat{k}^2$ is the square of the $d-4$-dimensional parts of the neutrino and gluon momenta,
which are the same in the adopted frame. It is thus the only $d-4$-dimensional invariant in the 
calculation~\cite{Gordon:1993qc,Jager:2002xm}.
Note that the $z$-integral cancels against the Beta function in the last line of~(\ref{ps3})
for all terms in the squared matrix element that have no dependence on $\hat{k}^2$.

Since the lepton pair is produced via an intermediate $W$ boson, a 
propagator with the momentum $p_\ell+p_{\nu_\ell}$ of the $W$ boson appears in the amplitude 
for the process. As a result, the squared matrix element $|{\cal {M}}|^2$ contains the overall factor
\be\label{s12prop}
\frac{1}{(s_{12}-M_W^2)^2+\Gamma_W^2 M_W^2}\;,
\ee
with the leptons' pair mass squared:
\be
s_{12}\equiv(p_\ell+p_{\nu_\ell})^2\;.
\ee
$s_{12}$ is a function of the angles $\theta_1$ and $\theta_2$. 
Since the neutrino is not observed, the propagator will be subject to integration over the 
phase space. We write it in the following way:
\ba\label{eq:pfmassive}
&&\hspace*{-1cm} \frac{1}{(s_{12}-M_W^2)^2+g^2} \nonumber \\[2mm]
&&\hspace*{-4mm}=\, \frac{1}{2i g}\left( \frac{1}{s_{12}-M_{W}^2-ig} -\frac{1}{s_{12}-M_{W}^2+ig} \right),
\ea
where $g\equiv\Gamma_W M_W$. After this partial fractioning, there are only terms in $|{\cal {M}}|^2$
with at most one power of $s_{12}$ in the denominator, either $1/(s_{12}-M_W^2-ig)$ or $1/(s_{12}-M_W^2+ig)$.
They are usually accompanied by other Mandelstam variables that also depend on $\theta_1$ and $\theta_2$. 
The ensuing terms may be readily integrated using the integrals
\ba\label{eq:integral2}
&&\hspace*{-8mm}I^{(k,n)}  =  \int_{0}^{\pi}d\theta_{1}\int_0^\pi d\theta_{2} \,
\sin^{1-2\varepsilon}{\theta_{1}}\sin^{-2\varepsilon}{\theta_{2}}  \nonumber \\[2mm]
&&\hspace*{-6mm} \times\frac{1}{(a+b\cos{\theta_{1}})^{k}(A+B\cos \theta_{1}+C\sin \theta_{1} \cos \theta_{2})^n} 
\ea
tabulated in the Appendix of Ref.~\cite{Beenakker:1988bq}. The results contain logarithms of various 
complex arguments which may be combined to produce manifestly real results. This procedure
is rather tedious; we have performed numerous numerical checks to ensure its correctness. 
For terms with dependence on $\hat{k}^2$ the $z$ integration in~(\ref{ps3}) is still trivial. The
result may then be further integrated using~(\ref{eq:integral2}). The additional power of 
$\sin^2\theta_1 \sin^2\theta_2$ arising from the $z$-integral can be easily accommodated
by shifting $\varepsilon\to \varepsilon-1$ in~(\ref{eq:integral2}). 

After integration over phase space the result for the real-gluon emission contribution contains 
singularities in $1/\varepsilon$. These occur whenever we have a term in $|{\cal {M}}|^2$ with at 
least a factor of  $1/t_3$ or $1/u_3$, where
\be
t_3=(p_q-p_g)^2,\;\; u_3=(p_{\bar{q}'}-p_g)^2 \, .
\ee
The poles arise when the gluon becomes collinear with the incoming particles, and/or when it 
becomes soft. The collinear singularities arise directly in the angular integrations.
A soft singularity is equivalent to the invariant mass squared of the two unobserved particles
becoming small, \textit{i.e.} $s_{23}=sv(1-w)\rightarrow 0$, or equivalently $w\rightarrow1$.
To make also the soft divergences manifest, we use the standard expansion 
\ba
(1-w)^{-1-\varepsilon}&=&-\frac{1}{\varepsilon} \delta(1-w)+\frac{1}{(1-w)_+}\nn\\[2mm]
&&-\varepsilon \left(\frac{\log(1-w)}{1-w}\right)_+ +\mathcal{O}(\varepsilon^2) \;,
\ea
where the ``plus'' distributions are defined as usual by
\be
\int_0^1dw\,f(w) [g(w)]_+=\int_0^1dw\,[f(w)-f(1)]\,g(w)\;.
\ee
The final expression contains quadratic ($1/\varepsilon^2$) poles as well as single ($1/\varepsilon$) ones.
We note that due to the finite width $\Gamma_W$ of the $W$ boson, final-state singularities never occur.

The NLO contributions associated with $qg\to \ell \nu q'$ scattering at NLO (Fig.~\ref{fig1_w1}(d)) can 
be integrated in the same way as described above. They develop only single poles in $1/\varepsilon$ 
since soft singularities are absent here.
\begin{figure}[t]
\vspace{0.2cm}
\includegraphics[width=0.48\textwidth]{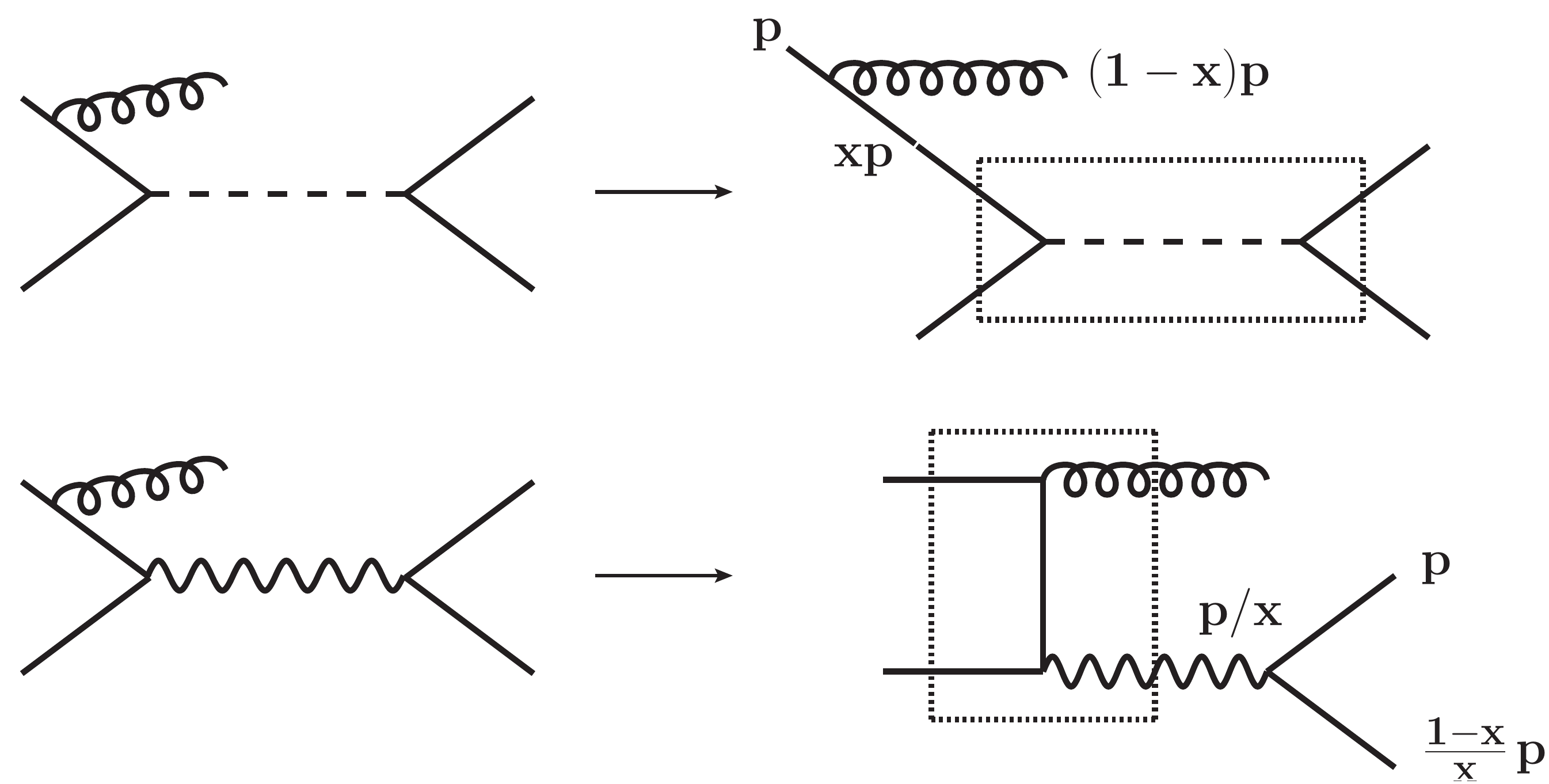}
\caption{\label{fig2_w2}\sf Upper row: Representative initial-state collinear 
contribution for the $q\bar{q}'$ channel. Lower row: 
Factorization of the final-state collinear singularity which is necessary for the process with an intermediate photon.}
\end{figure}

\subsection{Virtual correction and factorization of collinear singularities}

At NLO, the interference of the virtual diagrams (see for example Fig.~\ref{fig1_w1}(b)) with the Born diagram 
contributes. As may be inferred from~\cite{Altarelli:1979ub,Buras:1989xd}, the first-order virtual corrections 
only modify the basic $q\bar{q}'W$ vertex by a multiplicative factor of the form $1+{\cal O}(\alpha_s)$. 
Therefore, when computing the interference with the Born diagram,  the result will be twice the Born cross
section multiplied by this factor. In our notation, we have from~\cite{Altarelli:1979ub}:
\ba
\frac{d\hat{\sigma}_{q\bar{q}'}^{(1),\mathrm{virt}}}{dvdw}&=&
C_F \frac{d\hat{\sigma}_{q\bar{q}'}^{(0),\varepsilon}}{dv}\delta(1-w) \,
\left(-\frac{2}{\varepsilon^2}-\frac{3}{\varepsilon}-8+\pi^2\right)\nonumber \\[2mm]
&\times& \left(\frac{4\pi\mu^2}{s} \right)^\varepsilon \frac{\Gamma(1+\varepsilon)\,\Gamma^2(1-\varepsilon)}
{\Gamma(1-2\varepsilon)}\;,
\ea
where $C_F=4/3$. It is important to take into account here that the Born cross section is to be 
computed in $4-2 \varepsilon$ dimensions, where it is given by
\ba\label{bornn}
\frac{d\hat{\sigma}_{q\bar{q}'}^{(0),\varepsilon}}{dv}& = 
& \frac{|U_{qq'}|^2s}{8\pi N_c}\left(\frac{G_F M_W^2}{\sqrt{2}} \right)^2
 \left(\frac{4\pi}{s}\right)^\varepsilon \frac{(v(1-v))^{-\varepsilon}}{\Gamma(1-\varepsilon)}
\nonumber \\[2mm]
&\times&  \frac{C_1(1-v)^2+C_2 v^2-C_3\,\varepsilon}{(s-M_W^2)^2+\Gamma^2 M_W^2} \;.
\ea
Compared to the four-dimensional expression~(\ref{eq:borncross}) a new combination of the vector and 
axial vertex factors appears:
\be
C_3 = (a_\ell-v_\ell)^2 (a_q-v_q)^2\;.
\ee
As it turns out, this combination appears also in the real-emission contribution and in the factorization
subtraction discussed below, in such a way that the final result for the NLO correction only contains the combinations
$C_1$ and $C_2$ given in~(\ref{C12}). We furthermore note that the {\it spin-dependent} Born cross
section in $4-2 \varepsilon$ dimensions with an incoming polarized quark, $d\Delta \hat{\sigma}_{q\bar{q}'}^{(0),
\varepsilon}/dv$, is the negative of $d\Delta \hat{\sigma}_{q\bar{q}'}^{(0),\varepsilon}/dv$ in~(\ref{bornn}), 
but with $C_3=0$. This violation at order ${\cal O}(\varepsilon)$ of the relations in~(\ref{crsecrel}) and hence 
of helicity conservation is typical of intermediate results in the HVBM scheme~\cite{Gordon:1993qc,Jager:2002xm}. 

Adding the real and virtual contributions, the double poles in $\varepsilon$ cancel. We are left with single 
poles associated with collinear gluon emission. According to the factorization theorem, these may be absorbed 
into the parton distribution functions by a suitable subtraction which we perform in the $\overline{\mathrm{MS}}$ 
scheme. This introduces dependence on a factorization scale $\mu_F$. In the upper row of Fig.~\ref{fig2_w2}, one 
of the two initial-state collinear situations for the $2\rightarrow 3$ $q\bar{q}'$ channel is shown. 
Here, the variable $x$ denotes the momentum fraction of the incoming quark after radiating a gluon. 
The required subtraction is of the form $\sim \frac{1}{\varepsilon}\, P_{qq}\otimes
d\hat{\sigma}_{q\bar{q}'}^{(0),\varepsilon}$, where $P_{qq}$ is a LO Altarelli-Parisi splitting 
function~\cite{Altarelli:1977zs} and $d\hat{\sigma}_{q\bar{q}'}^{(0),\varepsilon}$ again the Born cross section 
for the process $q\bar{q}'\rightarrow\ell\nu_{\ell}$ computed in $4-2\varepsilon$ dimensions. 
More precisely, in case of the contribution shown in the upper part of Fig.~\ref{fig2_w2}, in the 
unpolarized case, we have to subtract the term
\ba
\label{subtractepsilon}
\frac{1}{vs}\frac{d\hat{\sigma}_{q\bar{q}'}^{(1),\mathrm{fact}}}{dvdw}&=&\int_{0}^{1}dx \,   
\frac{d \hat{\sigma}_{q\bar{q}'}^{(0),\varepsilon}(xs,xt,u,\varepsilon)}{dv} \nn\\[2mm]
&\times& H_{qq}(x,\mu_{F}) \;\delta(x(s+t)+u)\;,
\ea
where the $\overline{\mathrm{MS}}$ scheme is defined by
\be
H_{qq}(x,\mu_{F})=\left(- \frac{1}{\varepsilon}+\gamma_{E}-\log{4 \pi} \right)  
\left( \frac{\mu_{F}^{2}}{s} \right)^{-\varepsilon} P_{qq}(x) \;,
\ee
with $\gamma_E$ the Euler constant and with
\be\label{pqq}
P_{qq}(x)=C_F\left[ \frac{1+x^2}{(1-x)_+}+\frac{3}{2}\delta(1-x)\right]\;.
\ee
Standard $\overline{\mathrm{MS}}$ factorization requires the splitting function to be 
computed in four dimensions. After the collinear subtractions have been performed, 
we end up with the final NLO result in the $\overline{\mathrm{MS}}$ scheme. 

If the incoming quark is polarized, the subtraction is similar, but with two crucial differences:
First, one needs the spin-dependent Born cross section in $4-2 \varepsilon$ dimensions,
given as discussed above by the negative of the unpolarized one in~(\ref{bornn}) but with $C_3=0$. 
In addition, as discussed in Refs.~\cite{Gordon:1993qc,Vogelsang:1996im,Jager:2002xm}, 
in order to correct for violation of helicity conservation in the HVBM scheme, one needs to
use the splitting function
\be
\Delta P_{qq}(x)=C_F\left[ \frac{1+x^2}{(1-x)_+}+\frac{3}{2}\delta(1-x)+4\varepsilon(1-x)\right]
\ee
in the factorization subtraction. With these differences taken into account, the final spin-dependent 
NLO partonic cross sections respect the relations in~(\ref{crsecrel}), as they should. 
 
As already mentioned, in the case of an exchanged $W$ or $Z$ boson one does not encounter any 
final-state singularities. Effectively, the widths of the bosons act as regulators here. 
On the other hand, for an intermediate photon -- which presents one of the backgrounds -- 
a final-state singularity would occur if the leptons were massless, when the photon goes 
on its mass shell. Keeping a finite lepton mass is well beyond the scope of this work and is
also not necessary since the pure-photon contribution is in any case rather small. Also, 
because of parity conservation, it is only present in the unpolarized cross section and not
in the single-spin one. The artificial singularity that one encounters in this channel for 
massless leptons may be avoided for instance by imposing a cut on the invariant mass of
the outgoing lepton pair~\cite{deFlorian:2010aa}, or it may be simply subtracted in, say, 
the $\overline{\mathrm{MS}}$ scheme. Effectively, the latter approach, which we adopt here, 
introduces a (QED) photon-to-lepton fragmentation function~\cite{Kang:2008wv}. The diagrammatic 
situation for the final-state collinear splitting is shown in the lower row of Fig.~\ref{fig2_w2}. 
The subtraction to be performed is given by
\ba
\label{subtractepsilonneu}
\frac{1}{sv}\frac{d\hat{\sigma}_{q\bar{q}'}^{(1),\mathrm{photon\,fact}}}{dvdw}&=&
-\int_{0}^{1}dx \, \frac{d \hat{\sigma}^{(0)}_{q\bar{q} \rightarrow \gamma g}(s,t/x,u/x,\varepsilon)}{dv} \nn\\
&&\times \,H_{\ell\gamma}(x,\mu_{F}) \;\delta\left(s+\frac{t+u}{x}\right),
\ea
where $d\hat{\sigma}_{q\bar{q}\rightarrow \gamma g}$ denotes the Born-level cross section for the process 
$q\bar{q}\rightarrow \gamma g$ in $d=4-2\varepsilon$ dimensions, and where
\be
H_{\ell\gamma}(x,\mu_{F})=\left(- \frac{1}{\varepsilon}+\gamma_{E}-\ln{4 \pi} \right) \enspace 
P_{\ell\gamma}(x) \left( \frac{s}{\mu_{F}^{2}} \right)^{\varepsilon}\;,
\ee
with $P_{\ell\gamma}(x)$ the appropriate $\gamma\to \ell$ splitting function. Including the
thus defined subtraction renders the full NLO cross section finite. We stress again that the
pure-photon contribution is small, except at large lepton rapidities. It can in fact be vetoed
experimentally because it is characterized by two charged leptons that almost coalesce. We also note
that the $\gamma Z$ interference contribution does not produce any final-state singularities
even for massless leptons. 

Finally, for $qg$ scattering, there are no virtual corrections at ${\cal O}(\alpha_s)$. To obtain the finite 
cross section for these partonic channels, one therefore only needs the appropriate subtractions for
the initial-state collinear singularities.

\subsection{Final results}\label{ssec:finalresults}

Our final analytical NLO expressions for the processes $q\bar{q}'\to \ell X$, $qg \to \ell X$ 
through $W$-boson exchange are presented in the Appendix. We briefly summarize a few features 
of the result for the $q\bar{q}'\to \ell X$ channel. First of all, it contains the usual distributions in 
$(1-w)$, which dominate the cross section at $w\to 1$. These multiply the Born cross section:
\ba\label{eq:distributionsNLO}
\frac{d^2\hat{\sigma}^{(1)}_{q\bar{q}'}}{dvdw} &\stackrel{w\to 1}{\approx}&
\frac{d\hat{\sigma}_{q\bar{q}'}^{(0)}}{dv}C_F
\left[ 8 \left(\frac{\log(1-w)}{1-w}\right)_+ - \frac{4\,A(v) }{(1-w)_+} \right. \nn \\[2mm]
&&\hspace*{1.55cm}+\,B(v) \delta(1-w)\Bigg]\;,
\ea
where the coefficients $A(v),B(v)$ may be found from Eq.~(\ref{finalqqb}) in the Appendix. 
The terms with ``plus'' distributions represent the well-known threshold logarithms for the process
that arise when the incoming partons have just sufficient energy to produce the observed
final state, so that any substantial gluon radiation is kinematically inhibited.

The other terms in the NLO result have a more complicated structure. The integration of
terms containing~(\ref{s12prop}) gives rise to three different types of denominators.
We write them by introducing the function
\be
P(z)\,\equiv\,\frac{zs^2}{(zs-M_W^2)^2+\Gamma_W^2M_W^2}\;.
\ee
We then encounter the terms
\be\label{eq:denominators}
P_i\equiv P(z_i)\;\;\;\;\;\;\;\;(i=1,2,3)\;,
\ee
where
\be\label{zdef1}
z_1=1\, ,\;\; z_2 = w\,,\;\; z_3 = \frac{1-v}{1-vw}\;.
\ee
Evidently, $P_1$ essentially just corresponds to the propagator in the Born cross section. 
The other two propagators are similar and reduce to $P_1$ in the limit $w\to 1$. 

In addition to the new propagators arising at NLO, we also find several logarithms of the 
propagator terms. The logarithms that occur are 
\ba\label{eq:logs1}
&&\log\left( \frac{(ws-M_W^2)^2+\Gamma_W^2 M_W^2}{M_W^4+\Gamma_W^2M_W^2} \right)\;,\nn\\[2mm]
&&\log\left( \frac{(\frac{1-v}{1-vw}s-M_W^2)^2+\Gamma_W^2 M_W^2}{M_W^4+\Gamma_W^2M_W^2} \right)\;,\nn\\[2mm]
&&\log\left( \frac{((1-v+vw)s-M_W^2)^2+\Gamma_W^2 M_W^2}{M_W^4+\Gamma_W^2M_W^2} \right).
\ea
As seen in Eq.~(\ref{defkandz}), they are accompanied by inverse tangent functions resulting from the 
imaginary parts of the arguments of the logarithms arising in phase space integration. All these terms are 
multiplied by simple functions of $v$ and $w$ and by one of the three types of propagators
given above. The result for the channel $q g\to \ell X$ does not contain threshold distributions but 
does have logarithms of the type in Eq.~(\ref{eq:logs1}); see the Appendix for further details.

\section{Phenomenological results \label{sec:Wpnumericalresults}}

\begin{figure}[t]
\hspace*{-1.3cm}
\includegraphics[width=0.46\textwidth,angle=90]{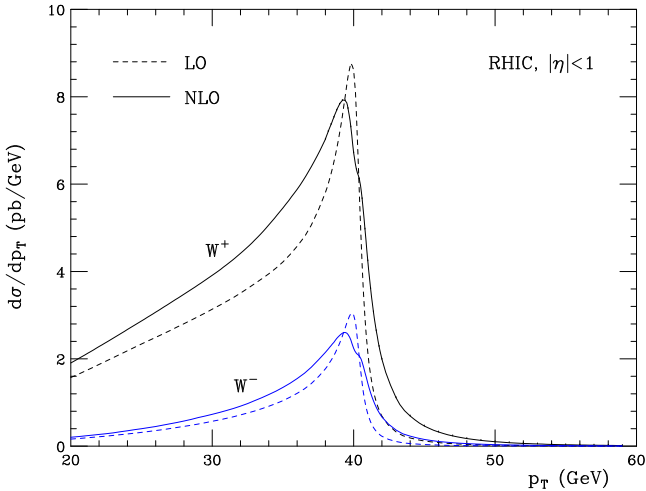}
\vspace*{-12mm}
\caption{\label{figrhic}\sf LO (dashed) and NLO (solid) cross sections at RHIC ($\sqrt{S}=510$~GeV) for
$\ell^+$ and $\ell^-$-production through $W^\pm$ boson exchange.}
\end{figure}

We start with the unpolarized cross section for $pp$ scattering at RHIC at $\sqrt{S}=510$~GeV. 
Figure~\ref{figrhic} shows our LO (dashed) and NLO (solid) results for the cross section 
$d\sigma/dp_T$ for $\ell^+$ and $\ell^-$ production through intermediate $W$ bosons. We
have integrated over $|\eta|\leq 1$ in the charged lepton's rapidity. We have used the NLO 
parton distributions of~\cite{Martin:2009iq} and the renormalization and factorization scales 
$\mu_R=\mu_F=p_T$. Our adopted values for the $W$ mass and width are $M_W=80.398$~GeV, 
$\Gamma_W=2.141$~GeV (later we will also use $M_Z=91.187$~GeV and $\Gamma_Z=2.49$~GeV 
for the $Z$ boson). 

\begin {figure*}[t]
\begin{center}
\vspace*{-6mm}
\hspace{-1.5cm}
\includegraphics[width=0.4\textwidth,clip=true,angle=90, trim=1cm 2cm 1cm 1cm ]{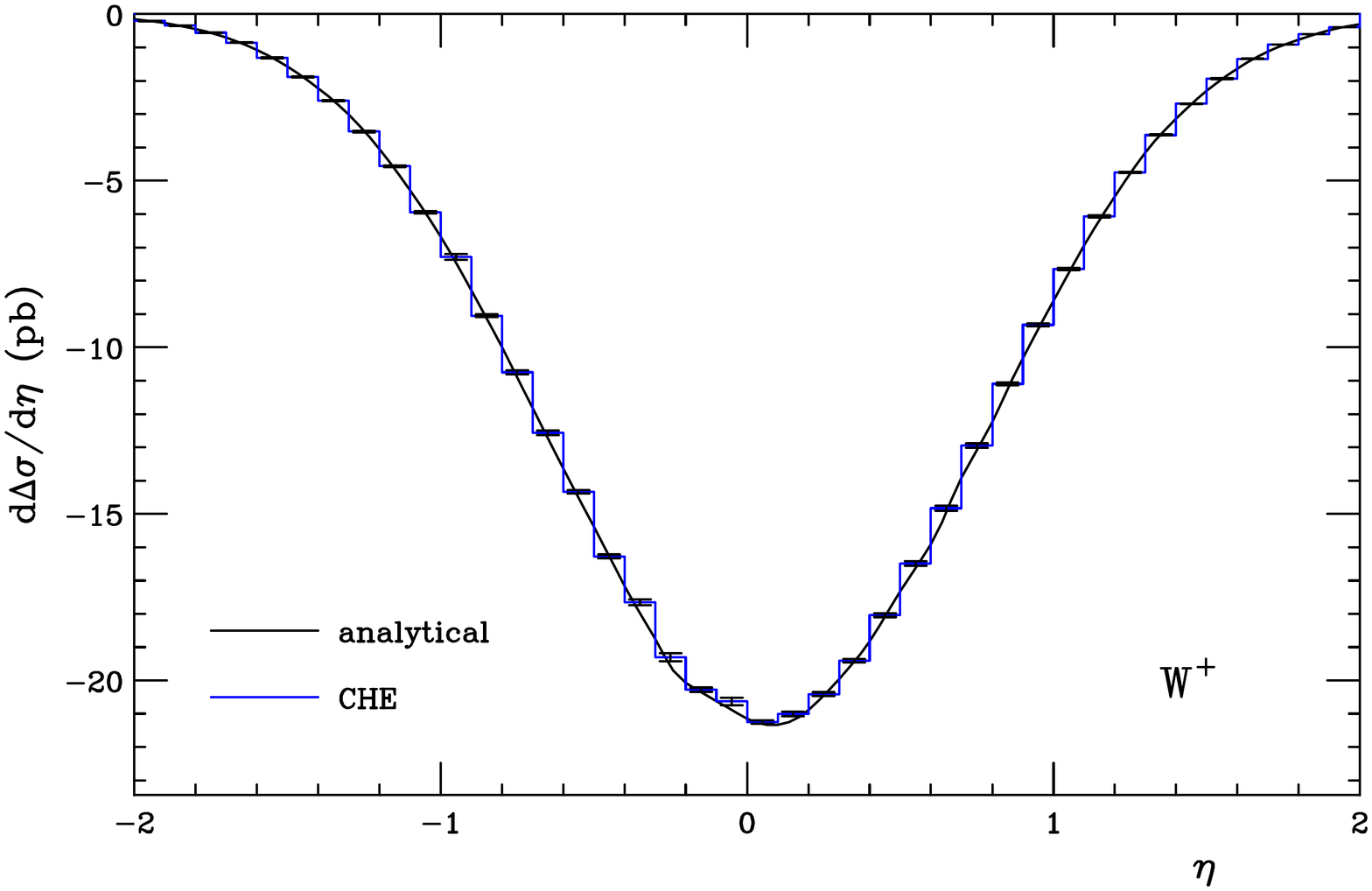} 
\hspace*{-0.2cm}
\includegraphics[width=0.4\textwidth,clip=true,angle=90,trim=1cm 2cm 1cm 1cm ]{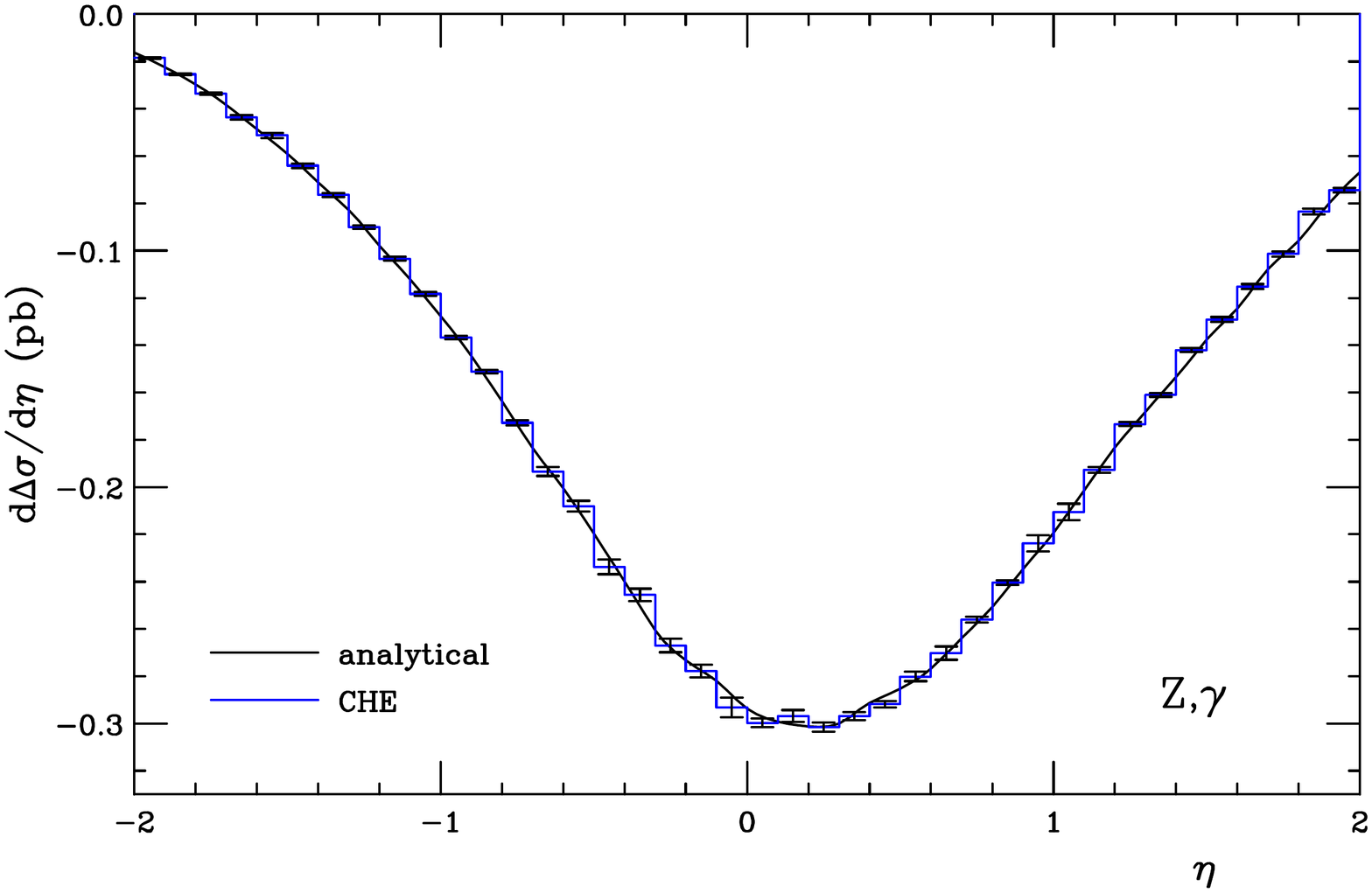} 
\hspace*{-1.cm}
\end{center}
\vspace*{-1.cm}
\caption{{\sf Comparison of our analytical results with the corresponding ones from 
CHE~\cite{deFlorian:2010aa} for the polarized cross sections $\Delta\sigma$ for $\ell^+$ production 
through $W^+$ decay (left) and through intermediate $Z$ or $\gamma$. We have considered here 
$pp$ collisions at $\sqrt{S}=500$~GeV and have integrated over $20\leq p_T\leq 60$~GeV.
As in~\cite{deFlorian:2010aa} the parton distributions have been chosen from 
Refs.~\cite{deFlorianpPDF,Martin:2002aw}.} 
\label{fig:deltasigma}}
\end{figure*}

Clearly, the NLO corrections are significant everywhere. They have 
moderate size below and around the Jacobian peak at $p_T\approx M_W/2$ and become very 
large well above the peak. A close inspection of the results in Fig.~\ref{figrhic} reveals a hint
of a ``shoulder'' in the NLO cross sections just above $p_T=M_W/2$. This shoulder is 
a true feature of the NLO results. It comes about in two ways: First, the $q\bar{q}'$ channel itself
has non-trivial structure here. Near $p_T=M_W/2$, there is a complicated interplay between positive
contributions by terms with distributions in $(1-w)$ (``plus distributions'' or $\delta$-function) 
in Eq.~(\ref{finalqqb}), and contributions by subleading terms in $(1-w)$, among them the 
terms involving the functions $J$ and $K$, which are negative around $p_T\sim M_W/2$ and 
become positive just below and above the Jacobian peak. 
This means that the $q\bar{q}'$ channel is sensitive to the exact 
mix of positive and negative contributions. Secondly, the $qg$ process 
makes a negative contribution below and around $p_T=M_W/2$ and then becomes positive.
This intricate interplay of the various contributions is also the reason why the height of the peak is 
reduced at NLO as compared to LO. We note that for increasing energy $\sqrt{S}$ the shoulder
becomes even more pronounced and in fact quickly turns into a double-peaked structure at NLO;
see also~\cite{Dittmaier:2014qza}. This at first sight surprising feature is a manifestation of the 
well-established fact~\cite{Smith:1983aa} that the region around the Jacobian peak cannot be 
controlled within a fixed-order calculation. Among other things, it is sensitive to small transverse 
momenta $q_T$ of the intermediate $W$ boson. There are large double-logarithmic 
corrections to the $q_T$-distribution of $W$ bosons at low $q_T$, which need to be taken 
into account to all orders if one wants to address this region~\cite{Balazs:1995nz}. 
Such a resummation is incorporated
in the RHICBOS code~\cite{Nadolsky:2003ga}. These issues become relevant for precision 
determinations of the mass of the $W$ boson from the lepton's $p_T$ spectrum near the
Jacobian peak~\cite{D0:2013jba}. For RHIC, they are not really relevant since, if one is interested in 
determining polarized parton distributions, there is no need to focus on the region around the Jacobian 
peak. Rather, it is advisable to integrate over a sizable range in $p_T$, so that the Jacobian peak
region constitutes only a rather small part of the cross section, and to  study the distribution of the charged 
lepton in rapidity. This is the strategy adopted by the RHIC experiments. We will therefore consider only
lepton rapidity distributions in the remainder of this paper. We plan to present a more detailed analysis 
of the region around the Jacobian peak in future work.

In order to check the validity of our analytical results and their numerical evaluation, we have
performed extensive comparisons to high-statistics runs of the NLO code CHE presented in 
Ref.~\cite{deFlorian:2010aa}, both for the unpolarized and for the polarized case. We have found 
excellent agreement. A representative example is given in Fig.~\ref{fig:deltasigma}, where
we show the spin-dependent cross sections for $\ell^+$ production at RHIC, through 
$W^+$ boson exchange (left) and for the background channels, $Z$-boson exchange and $\gamma Z$ 
interference (right; the pure-photon channel does not contribute to the spin-dependent cross section). 
Both our analytical (solid lines) and the CHE results (histograms) are shown. We have 
followed~\cite{deFlorian:2010aa} to use the polarized parton distributions of~\cite{deFlorianpPDF}
(referred to as DSSV08) and the unpolarized ones of~\cite{Martin:2002aw} which were also the
baseline set in the DSSV08 global analysis. Furthermore, the figure is for $\sqrt{S}=500$ GeV, and 
the transverse momentum of the observed charged lepton has been integrated over the 
range of $20<p_T<60$ GeV. As in~\cite{deFlorian:2010aa} we have chosen the renormalization 
and factorization scales as $\mu_R=\mu_F\equiv\mu=\sqrt{p_T^2+M_W^2}/2$ and assumed $n_f=4$ active 
quark flavors. In Fig.~\ref{fig:deltasigma} the error bars of the results shown for 
CHE correspond to numerical integration uncertainties. The uncertainties in our new numerical 
calculation are smaller than the widths of the lines. Since our results are largely analytical whereas 
the code of~\cite{deFlorian:2010aa} is based on a standard Monte-Carlo integration with numerical 
cancelation of singularities, our new code produces the results shown in about two orders of magnitude 
less time. Of course, Monte-Carlo based codes are more flexible in general, allowing the implementation of
various additional kinematical cuts and observables if necessary. 

\begin {figure*}[t]
\begin{center}
\vspace*{-6mm}
\hspace{-1.4cm}
\includegraphics[width=0.4\textwidth,clip=true,angle=90, trim=1cm 2cm 1cm 1cm ]{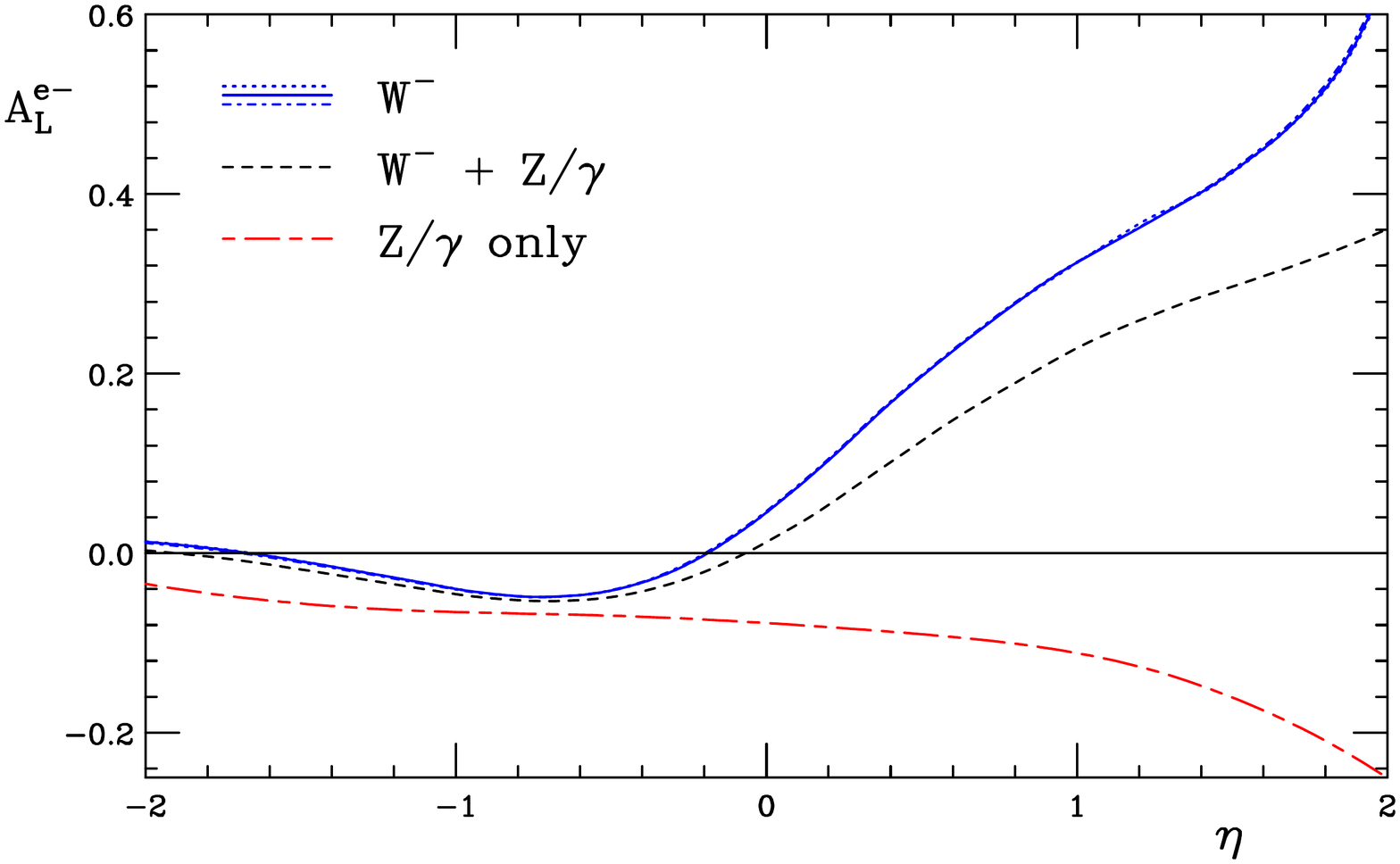} 
\hspace*{-0.2cm}
\includegraphics[width=0.4\textwidth,clip=true,angle=90,trim=1cm 2cm 1cm 1cm ]{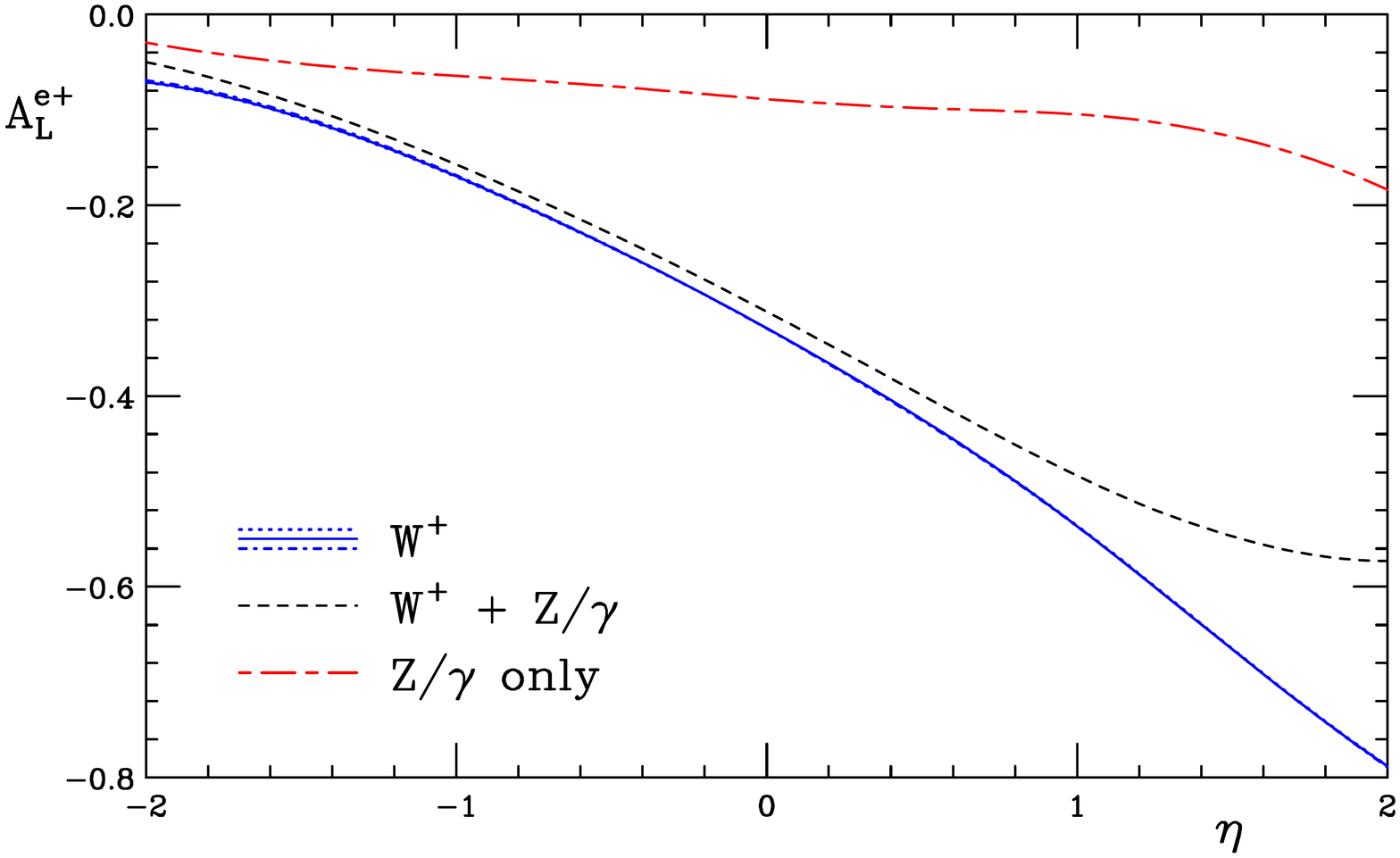} 
\hspace*{-1.cm}
\end{center}
\vspace*{-1.cm}
\caption{{\sf Single-spin asymmetries $A_L^{\ell^{\pm}}$ for negatively (left) and positively
(right) charged leptons as functions of rapidity at $\sqrt{S}=510$~GeV. We have integrated
over the range  $p_T\geq 30$ GeV. The solid lines show the results when the lepton
originates exclusively from $W$ bosons, at scale $\mu=M_W/2$. The dotted and dot-dashed 
lines correspond to the scale choices $\mu=p_T$ and $\mu=M_W$, respectively (note that
the lines for the various scales are almost indistinguishable).
For the dashed lines the background from exchanged $Z$ bosons and photons
has been included, using the scale $\mu=M_W/2$. Finally, the long-dashed lines show
the spin asymmetries for $Z$ bosons and photons alone, without the $W$-boson contributions,
this time for $25<p_T<50$ GeV.
We have used the DSSV14 polarized parton distributions~\cite{deFlorianpPDF} and the unpolarized 
ones of~\cite{Martin:2009iq}. } \label{as_pm} }
\end{figure*}

We now turn to the spin asymmetries $A_L$ which are the quantities of primary
interest in RHIC's $W$ physics program. Figure~\ref{as_pm} shows our NLO results 
at $\sqrt{S}=510$~GeV as functions of $\eta$. The cross sections have been integrated 
over $p_T\geq 30$~GeV, as appropriate for comparison to the PHENIX
data~\cite{phenix,Adare:2010xa}. We have now used the new set of polarized parton 
distributions of Ref.~\cite{deFlorian:2014yva} (referred to as DSSV14). This set primarily 
contains updated information on the nucleon's spin-dependent gluon distribution, which is less
relevant for weak boson production. However, it is also based on new results from inclusive and 
semi-inclusive lepton scattering~\cite{Alekseev:2010hc}, so that it offers new information on 
the quark and antiquark helicity distributions as well. We use the unpolarized parton distributions 
of~\cite{Martin:2009iq}. The solid lines in the figure show our results for charged-lepton production 
via $W$ decay for the scale choice $\mu=M_W/2$, while the dotted and dot-dashed lines correspond to the
choices $\mu=p_T$ and $\mu=M_W$, respectively. One can see that the scale dependence of the asymmetries
is extremely weak, which is one of the reasons why $W$ boson production at RHIC is an excellent
and theoretically well-controlled probe of nucleon spin structure. In Fig.~\ref{as_pm} we also 
investigate the impact of the ``background'' presented by $Z$ and $\gamma$ exchange. The dashed
lines show the NLO results for the scale $\mu=M_W/2$, now including the $Z$ and photon 
contributions. As is known from previous studies~\cite{Nadolsky:2003ga,deFlorian:2010aa}, the 
background channels dilute the spin asymmetries somewhat, which is mostly due to the increase of the 
unpolarized cross section in the denominator of the asymmetry. We note that the STAR experiment
at RHIC is able to identify and subtract this background, using data as well as Monte-Carlo estimates, 
so that the data can be directly compared to calculations based on only intermediate $W$ bosons. 
For comparisons to PHENIX data, the $Z/\gamma$ background needs to be included. Figure~\ref{as_pm} 
also shows the spin asymmetries for $Z$ and $\gamma$ exchange alone, in this case integrated over 
$25<p_T<50$ GeV corresponding to conditions in STAR~\cite{Adamczyk:2014xyw}.

\begin{figure}[t]
\vspace*{-1cm}
\hspace*{-0.8cm}
\includegraphics[width=0.44\textwidth,angle=90]{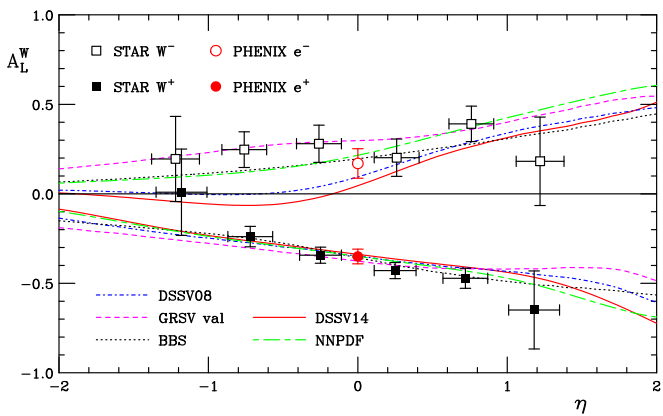}
\vspace*{-1.5cm}
\caption{\label{fig:STAR} {\sf Comparisons of NLO results for $A_L^{W^\pm}$ for various sets
of helicity parton distributions~\cite{deFlorian:2014yva,deFlorianpPDF,Nocera:2014gqa,Gluck:2000dy,Bourrely:2001du}
to the STAR data~\cite{Adamczyk:2014xyw} taken at $\sqrt{S}=510$ GeV and to the PHENIX
mid-rapidity points for electrons/positrons with $|\eta|\leq 0.35$~\cite{phenix}. The cut $25<p_T<50$ GeV has been applied on the lepton's 
transverse momentum. Note that the PHENIX points are for $p_T>30$ GeV and includes the contributions from photons 
and $Z$ bosons. We have chosen the scales $\mu_R=\mu_F=M_W/2$.}}
\end{figure}

Using our new NLO code, we finally compare in Fig.~\ref{fig:STAR} 
the results for various sets of spin-dependent parton distributions to
the published STAR~\cite{Adamczyk:2014xyw} and PHENIX~\cite{phenix} spin asymmetry data taken at $\sqrt{S}=510$ GeV.
The STAR $A_L^W$ data have been presented for various $\eta$, sampled over the range $25<p_T<50$ GeV of lepton transverse 
momenta, and our theoretical results shown are adapted to these conditions. We note that for PHENIX the cut on 
transverse momentum is different, $p_T>30$ GeV, and the asymmetry is for electrons or positrons and hence 
includes the contributions from photons and $Z$ bosons, as we just discussed. These are, however,
relatively small effects (see Fig.~\ref{as_pm}), so we show the PHENIX data point along with our results and the STAR points. 
In view of the results shown in Fig.~\ref{as_pm} the scale choice hardly matters; we
use $\mu_R=\mu_F=M_W/2$. The sets of spin-dependent parton distributions we use are 
from~\cite{deFlorianpPDF,deFlorian:2014yva} (DSSV08 and DSSV14), from~\cite{Nocera:2014gqa} (NNPDFpol1.1),
as well as the ``statistical'' parton distributions of~\cite{Bourrely:2001du,Bourrely:2013qfa} 
and a much earlier set~\cite{Gluck:2000dy} known as the ``GRSV valence scenario''. From the 
figure we draw the following observations:
\begin{itemize}[leftmargin=*]
\item all sets describe the $W^+$ asymmetry data rather well. The main reason for this is that the spin asymmetry
is largely driven by the polarized up quark distribution which is relatively well constrained by DIS data and
hence similar in all sets.  
\item among the various sets, NNPDFpol1.1 is the only one for which the STAR data were already included in the
analysis, constraining the light sea quark helicity distributions. As a result, the data are quite well described
by the set, especially when one includes the corresponding uncertainty estimates~\cite{Nocera:2014gqa} that 
we do not show here. Note, however, that information from semi-inclusive lepton scattering is not included
in the NNPDFpol1.1 set. 
\item at $\eta\leq 0$, the two DSSV sets show $W^-$ asymmetries that are below the data. Since
the DSSV14 set contains the latest information available from (semi-inclusive) DIS, this hints at the
interesting possibility of a tension between the DIS and RHIC data, the latter favoring a larger 
$\Delta\bar{u}$ distribution (see also the discussion in~\cite{Nocera:2014gqa}). It has to be emphasized,
however, that we do not display here any uncertainties for the DSSV set; as shown 
in~\cite{deFlorianpPDF,Adamczyk:2014xyw}, the main DSSV08 uncertainty band is such that it just about touches
the lower end of the error bars of the data points. In this sense, it is premature to draw any conclusions regarding
such a tension. Clearly, it will be interesting to follow up on this issue in the context of a new global analysis,
especially when additional experimental information becomes available.
\item in the framework of the statistical parton distributions, the helicity distributions are obtained
along with the unpolarized ones and depend on only very few parameters to be determined from data.
As one can see from Fig.~\ref{fig:STAR} (and as discussed in~\cite{Bourrely:2013qfa}), the model
describes the RHIC data quite well. 
\item the GRSV valence scenario of~\cite{Gluck:2000dy} describes the $W^-$ asymmetry data 
strikingly well. The main distinctive features for this set are assumptions about the breaking of
SU(3) in the relations between nucleon spin structure and hyperon $\beta$-decays, and the 
ansatz~\cite{Gluck:2000ch}
\be
\frac{\Delta\bar d(x,Q_0^2)}{\Delta\bar u(x,Q_0^2)}\,=\,\frac{\Delta u(x,Q_0^2)}{\Delta d(x,Q_0^2)}
\ee
at a low initial scale $Q_0$. Since $\Delta u$ and $\Delta d$ are known to have opposite sign,
the latter ansatz forces the ratio $\Delta\bar d/\Delta\bar u$ to be negative. This requirement,
along with the condition $\Delta\bar u+\Delta\bar d<0$ imposed by the DIS data and the 
assumptions about SU(3)-breaking, is realized in this model by a fairly large positive $\Delta\bar u$
distribution and a negative (and even larger in absolute value) $\Delta\bar d$ one. Evidently,
the STAR data prefer such a sizable positive $\Delta\bar u$. We note that one of the sets of
Ref.~\cite{deFlorian:2005mw} has a similar $\Delta\bar u$ distribution and hence describes
the $W^-$ asymmetry data similarly well~\cite{deFlorian:2010aa}. It will be interesting to see
whether also the large negative $\Delta\bar d$ of~\cite{Gluck:2000dy} is realized; unfortunately, 
the $\Delta\bar d$ contribution to the $W^+$ asymmetry is typically overwhelmed by the $\Delta u$ one. 
Note that a negative $\Delta\bar d$ pulls the $W^+$ asymmetry to more negative values  
(see~(\ref{eq:w+lo}) in the introduction), which may explain why the GRSV valence 
scenario shows the most negative asymmetry of all the sets at $\eta\leq 0$. Needless to say 
that the GRSV valence scenario has not been confronted with the latest (semi-inclusive) DIS data.
\end{itemize}

\section{Conclusions \label{sec:Wpconclusions}}

We have presented a new analytical NLO calculation of the partonic cross sections for single-inclusive 
lepton production at RHIC, when the lepton originates from the decay of an intermediate 
electroweak boson, especially a $W$ boson. Our numerical code based on analytical phase space 
integration is much faster than existing Monte-Carlo integration based codes. In this way,
we hope that our code will be a valuable tool for future global analyses of the proton's helicity parton 
distributions that include the new high-precision data for $A_L^W$ asymmetries obtained at RHIC.
Our results may also be useful to obtain insights into the analytical structure of the partonic cross sections, 
for example in terms of their threshold logarithms or their behavior in the vicinity of the Jacobian peak. 

We have also presented new comparisons of the latest RHIC data with the NLO predictions for some of 
the sets of polarized parton distributions available in the literature. In line with observations in
the earlier literature we have found that the data prefer a rather sizable positive $\Delta \bar{u}$ helicity
distribution in the proton.

\section*{Acknowledgments}

We are grateful to Abhay Deshpande, Daniel~de~Florian, Ciprian Gal, Alexander Huss, Jacques Soffer, 
Marco Stratmann, and Bernd Surrow for helpful discussions and communications. 

\section*{Appendix}

\renewcommand{\theequation}{A.\arabic{equation}}
\setcounter{equation}{0} 

In this Appendix, we present some of our explicit NLO results. We first consider the $q\bar{q}'$ channel when
an intermediate $W^-$ boson is produced (for example through $d\bar{u}$ scattering), for which effectively 
$C_1=0, C_2=8$ in~(\ref{C12}) (see discussion after Eq.~(\ref{eq:borncross})). We define the functions
\ba\label{defkandz}
K(z)&\equiv& \arctan\left( \frac{\Gamma_W M_W}{zs-M_W^2}\right) +\pi \,\Theta(M_W^2-zs)\;,\nn\\[2mm]
J(z)&\equiv&\log\left[\frac{(zs-M_W^2)^2+\Gamma_W^2 M_W^2}{M_W^4+\Gamma_W^2M_W^2}\right]-
\frac{2M_W}{\Gamma_W}K(z)\;,\nn\\
\ea
with the usual (Heaviside) step function. In addition to the values $z_1=1\, ,
z_2= w\,,z_3= (1-v)/(1-vw)$ of Eq.~(\ref{zdef1}), we introduce
\be
z_0=0\, ,\;\;\;z_4=1-v+v w\;,
\ee
and we set
\be
J_i\,\equiv\,J(z_i)\,,\;\;\;\;K_i\,\equiv\,K(z_i)\,.
\ee
We then find for production of a $W^-$:
\begin{widetext}
\ba\label{finalqqb}
\frac{s\,d^2\hat{\sigma}^{(1)}_{q\bar{q}'}}{dvdw}& =&
\frac{|U_{qq'}|^2}{\pi N_c}\left(\frac{G_F M_W^2}{\sqrt{2}}\right)^2C_F\,
\left[v^2 P_1 \left[ 2(1+w^2) \left(\frac{\log(1-w)}{1-w}\right)_+ 
- 2\,\log(1-v w)\,\frac{P_{qq}(w)}{C_F} \right.\right.\nn \\[2mm]
&&\hspace*{-1cm}+\,\left(\pi^2-8+\left(\frac{3}{2}+2\log(1-v)\right)
\log\frac{1-v}{v}\right) \delta(1-w) +\frac{1+w^2}{1-w}\left(
J_0-J_2-J_3+J_4+\kappa(K_0-K_2-K_3+K_4)\right)\Bigg]\nn\\[2mm]
&&\hspace*{-1cm}-\;\frac{v}{2}\left(\frac{J_0-2J_3+J_4}{1-vw}-\frac{J_0-J_4}{1-v+vw}\right)+ 
v^2\left\{P_2\left[ (1+w^2) \left(\frac{\log(1-w)}{1-w}\right)_+ 
-\frac{P_{qq}(w)}{C_F}\log\left(\frac{\mu_F^2}{vs}\right)+1-w\right.\right.\nn\\[2mm]
&&\hspace*{-1cm}-\,\left.\left.\left.\!\!\frac{1}{2}\,\frac{1+w^2}{1-w} 
\left(J_0-2J_2+J_4+\frac{\kappa}{w}(K_0-2 K_2+K_4)\right)\right]\right\}+
\frac{v^3w^2}{1-vw}\Bigg\{v\to 1-v w,w\to \frac{1-v}{1-v w}\Bigg\}\right],
\ea
\end{widetext}
with the splitting function $P_{qq}$ of Eq.~(\ref{pqq}), and with
\be
\kappa\equiv\frac{2M_W(\Gamma_W^2+M_W^2)}{\Gamma_W s}\;.
\ee
Note that despite appearance the expression is perfectly well regularized at $w=1$.

By applying crossing one obtains the corresponding cross section for $\bar{q}'q\to W^-g$. 
Crossing is achieved by changing $v\to 1-vw$, $w\to (1-v)/(1-vw)$ and multiplying the result 
by the Jacobian $v/(1-vw)$. We do not give the crossed result explicitly here. 

Writing the NLO partonic $q\bar{q}'$ cross section for general $C_1$ and $C_2$ in the form 
\beq\label{C12new}
C_1 \,d\hat{\sigma}^{(1)}_1 + C_2 \,d\hat{\sigma}^{(1)}_2\,,
\eeq
we find that $d\hat{\sigma}^{(1)}_2=[d\hat{\sigma}^{(1)}_1]_{\mathrm{crossed}}$. Since
the result for $W^+$ production is obtained in our calculations by setting $C_1=8, C_2=0$
(see Sec.~\ref{ssec:Born}), we thus have
\ba
d\hat{\sigma}^{(1)}_{q\bar{q}'\to W^+g} &=&8 d\hat{\sigma}^{(1)}_1 \;=\;
d\hat{\sigma}^{(1)}_{\bar{q}q'\to W^-g}\,,\nn\\[2mm]
d\hat{\sigma}^{(1)}_{\bar{q}'q\to W^+g}&=&8 \left[d\hat{\sigma}^{(1)}_1\right] _{\mathrm{crossed}}\;=\;
d\hat{\sigma}^{(1)}_{q'\bar{q}\to W^-g} .\;\;
\ea
We remind the reader that the $W^\pm$ cross section for a polarized incoming quark differs just by a 
sign from the corresponding unpolarized one (see Eq.~(\ref{crsecrel})) while that for
an incoming polarized antiquark involves no sign change. The cross sections for intermediate $Z$ 
bosons may be constructed from~(\ref{C12new}), using~(\ref{finalqqb}) 
and its crossed variant and inserting the appropriate coupling factors $C_1$ and $C_2$ in each case.

Secondly, we also present the result for the channel $g \bar q\to W^-\bar{q}'$ in the unpolarized and the polarized case:
\begin{widetext}
\ba
\frac{s\,d^2\hat\sigma^{(1)}_{g \bar q}}{dvdw} & = &\frac{T_R|U_{qq'}|^2}{\pi N_c}\left(\frac{G_F M_W^2}{\sqrt{2}}\right)^2 
v^2P_2\,\Bigg\{ 2\, (1-w) w - \, P_{qg}(w) \nn \\[2mm]
&& \times \left[J_0-2J_2+J_4 + \frac{\kappa}{w}(K_0-2K_2+K_4)+ 2\log\left(\frac{\mu_F^2}{v(1-w)s}\right) \right] \Bigg\} \, ,
\nn \\[2mm]
\frac{s\,d^2\Delta\hat\sigma^{(1)}_{g \bar q}}{dvdw} & = &-\frac{T_R|U_{qq'}|^2}{\pi N_c}\left(\frac{G_F M_W^2}{\sqrt{2}}\right)^2 
v^2P_2\,\Bigg\{ 2\, (1-w) - \, \Delta P_{qg}(w) \nn \\[2mm]
&& \times \left[J_0-2J_2+J_4 + \frac{\kappa}{w}(K_0-2K_2+K_4)+ 2\log\left(\frac{\mu_F^2}{v(1-w)s}\right) \right] \Bigg\} \, ,
\ea
\end{widetext}
where $T_R=1/2$ and
\ba
P_{qg}(x)&=&\frac{1}{2}\left(x^2 +(1-x)^2\right)\,,\nn\\[2mm]
\Delta P_{qg}(x)&=&\frac{1}{2}\left(2x-1\right)\,.
\ea
We note that the terms in square brackets have a similar structure as the penultimate one in~(\ref{finalqqb}).
Finally, for $qg\to W^-q'$ we find
\begin{widetext}
\ba
\frac{s\,d^2\hat{\sigma}^{(1)}_{qg}}{dvdw}& =&
\frac{T_R|U_{qq'}|^2}{\pi N_c}\left(\frac{G_F M_W^2}{\sqrt{2}}\right)^2\,
\left[ \frac{v}{1-v w} \left\{ \frac{2 M_W^2}{s} \left(J_0-2J_3+J_4 + \tilde{\kappa}
(K_0-2K_3+K_4)\right)\right.\right.\nn\\[2mm]
&+&\left.\left.P_3\,v^2w^2
\left[2(1-\tilde{w})\tilde{w}-P_{qg}(\tilde{w}) \, \left(J_0-2J_3+J_4 + 
\frac{\kappa}{\tilde{w}}(K_0-2K_3+K_4)+ 2\log\left(\frac{\mu_F^2}{v(1-w)s}\right) \right)\right]
\right.\right.\nn\\[2mm]
&-&(J_0-2 J_3+J_4) \, \frac{1-v-vw+2v^2 w-v^2 w^2}{1-vw} -\frac{(1+vw)(1-2v+vw)}{1-vw}\bigg\}\nn\\[2mm]
&+&\frac{v}{(1-v+v w)^2}\left\{-\frac{M_W^2}{s}\,\left(J_0-J_4+\tilde{\kappa}(K_0-K_4)\right)\,
\frac{1-3v+2 v^2+4 vw-3 v^2w+v^2 w^2}{1-v+vw}\right.\nn\\[2mm]
&+&\frac{1}{2}(J_0-J_4)(1-v)(1-2v+2 vw)-v(1-v-2w+vw)\bigg\}\bigg]\,,
\ea
\end{widetext}
where
\be
\tilde{w}\equiv\frac{1-v}{1-vw}\,=\,z_3\,,
\eeq
and
\beq
\tilde{\kappa}\,\equiv\,\frac{\Gamma_W^2+M_W^2}{\Gamma_W M_W}\;.
\eeq
The corresponding spin-dependent cross section for an incoming polarized quark again just differs by
a sign; see~(\ref{crsecrel}). 



\begin{thebibliography}{99}

\bibitem{Aschenauer:2015eha} 
  E.~C.~Aschenauer {\it et al.},
  arXiv:1501.01220 [nucl-ex].

\bibitem{Adamczyk:2014xyw} 
  L.~Adamczyk {\it et al.}  [STAR Collaboration],
  Phys.\ Rev.\ Lett.\  {\bf 113}, 072301 (2014)
  [arXiv:1404.6880 [nucl-ex]].

\bibitem{phenix} A.~Adare {\it et al.}  [PHENIX Collaboration],
  arXiv:1504.07451 [hep-ex].

\bibitem{Adare:2010xa} 
  A.~Adare {\it et al.}  [PHENIX Collaboration],
  Phys.\ Rev.\ Lett.\  {\bf 106}, 062001 (2011)
  [arXiv:1009.0505 [hep-ex]].

\bibitem{Aggarwal:2010vc} 
  M.~M.~Aggarwal {\it et al.}  [STAR Collaboration],
  Phys.\ Rev.\ Lett.\  {\bf 106}, 062002 (2011)
  [arXiv:1009.0326 [hep-ex]].

\bibitem{Bourrely:1993dd} 
  C.~Bourrely and J.~Soffer,
  Phys.\ Lett.\ B {\bf 314}, 132 (1993);
  Nucl.\ Phys.\ B {\bf 423}, 329 (1994)
  [hep-ph/9405250].

\bibitem{Bunce:2000uv} {\it see also:}
  G.~Bunce, N.~Saito, J.~Soffer and W.~Vogelsang,
  Ann.\ Rev.\ Nucl.\ Part.\ Sci.\  {\bf 50}, 525 (2000)
  [hep-ph/0007218].

\bibitem{Airapetian:2004zf} 
  A.~Airapetian {\it et al.}  [HERMES Collaboration],
  Phys.\ Rev.\ D {\bf 71}, 012003 (2005)
  [hep-ex/0407032].
  
\bibitem{Alekseev:2010hc} 
  M.~G.~Alekseev {\it et al.}  [COMPASS Collaboration],
  Phys.\ Lett.\ B {\bf 690}, 466 (2010)
  [arXiv:1001.4654 [hep-ex]];
  Phys.\ Lett.\ B {\bf 693}, 227 (2010)
  [arXiv:1007.4061 [hep-ex]].
    
\bibitem{deFlorian:2014yva} 
  D.~de Florian, R.~Sassot, M.~Stratmann and W.~Vogelsang,
  Phys.\ Rev.\ Lett.\  {\bf 113}, 012001 (2014)
  [arXiv:1404.4293 [hep-ph]].

\bibitem{deFlorianpPDF}
D.~de Florian, R.~Sassot, M.~Stratmann and W.~Vogelsang,
  Phys.\ Rev.\ Lett.\  {\bf 101}, 072001 (2008);
  Phys.\ Rev.\ D {\bf 80}, 034030 (2009).

\bibitem{Leader:2010rb} 
  E.~Leader, A.~V.~Sidorov and D.~B.~Stamenov,
  Phys.\ Rev.\ D {\bf 82}, 114018 (2010)
  [arXiv:1010.0574 [hep-ph]].

\bibitem{Kamal:1997fg} 
  B.~Kamal,
  Phys.\ Rev.\ D {\bf 57}, 6663 (1998)
  [hep-ph/9710374].

\bibitem{Gehrmann:1997ez} 
  T.~Gehrmann,
  Nucl.\ Phys.\ B {\bf 534}, 21 (1998)
  [hep-ph/9710508].

\bibitem{Weber:1993xm}   
  A.~Weber,
  Nucl.\ Phys.\ B {\bf 403}, 545 (1993).
  
\bibitem{Mukherjee:2006uu} 
  A.~Mukherjee and W.~Vogelsang,
  Phys.\ Rev.\ D {\bf 73}, 074005 (2006)
  [hep-ph/0601162].  
  
\bibitem{Nadolsky:2003ga} 
  P.~M.~Nadolsky and C.~P.~Yuan,
    Nucl.\ Phys.\ B {\bf 666}, 3 (2003)
  [hep-ph/0304001];
  Nucl.\ Phys.\ B {\bf 666}, 31 (2003)  [hep-ph/0304002].

\bibitem{deFlorian:2010aa} 
  D.~de Florian and W.~Vogelsang,
  Phys.\ Rev.\ D {\bf 81}, 094020 (2010)
  [arXiv:1003.4533 [hep-ph]].

\bibitem{vonArx:2011fz} 
  C.~von Arx and T.~Gehrmann,
  Phys.\ Lett.\ B {\bf 700}, 49 (2011)
  [arXiv:1103.1465 [hep-ph]].

\bibitem{Nocera:2014gqa}
  E.~R.~Nocera {\it et al.}  [NNPDF Collaboration],
  Nucl.\ Phys.\ B {\bf 887}, 276 (2014)
  [arXiv:1406.5539 [hep-ph]].

\bibitem{Aurenche:1980tp}
P.~Aurenche and J.~Lindfors,
Nucl.\ Phys.\ B {\bf 185}, 274 (1981).

\bibitem{Stratmann:2001pb} 
  M.~Stratmann and W.~Vogelsang,
  Phys.\ Rev.\ D {\bf 64}, 114007 (2001)
  [hep-ph/0107064].
    
\bibitem{Collins:1989gx} 
  J.~C.~Collins, D.~E.~Soper and G.~F.~Sterman,
  Adv.\ Ser.\ Direct.\ High Energy Phys.\  {\bf 5}, 1 (1988)
  [hep-ph/0409313].

\bibitem{'tHooft:1972fi}
G.~'t~Hooft and M.~Veltman,
Nucl.\ Phys.\ B {\bf 44}, 189 (1972).

\bibitem{Breitenlohner:1977hr}
P.~Breitenlohner and D.~Maison,
Commun.\ Math.\ Phys. {52}, 11 (1977).

\bibitem{Jamin:1991dp}
  M.~Jamin and M.~E.~Lautenbacher,
  Comput.\ Phys.\ Commun.\  {\bf 74}, 265 (1993).
  
\bibitem{Buras:1989xd}
  A.~J.~Buras and P.~H.~Weisz,
  Nucl.\ Phys.\ B {\bf 333}, 66 (1990).

\bibitem{Ciuchini:1993fk} 
  M.~Ciuchini, E.~Franco, L.~Reina and L.~Silvestrini,
  Nucl.\ Phys.\ B {\bf 421}, 41 (1994)
  [hep-ph/9311357].

\bibitem{Gordon:1993qc}
L.~E.~Gordon and W.~Vogelsang,
Phys.\ Rev.\ D {\bf 48}, 3136 (1993).

\bibitem{Vogelsang:1996im} 
W.~Vogelsang,
  Phys.\ Rev.\ D {\bf 54}, 2023 (1996)  [hep-ph/9512218];
  Nucl.\ Phys.\ B {\bf 475}, 47 (1996)  [hep-ph/9603366].
  
\bibitem{Jager:2002xm} 
  B.~J\"{a}ger, A.~Sch\"{a}fer, M.~Stratmann and W.~Vogelsang,
  Phys.\ Rev.\ D {\bf 67}, 054005 (2003)
  [hep-ph/0211007].

\bibitem{Goria:2011wa} 
  S.~Goria, G.~Passarino and D.~Rosco,
  Nucl.\ Phys.\ B {\bf 864}, 530 (2012)
  [arXiv:1112.5517 [hep-ph]].

\bibitem{Beenakker:1988bq} 
W.~Beenakker, H.~Kuijf, W.~van~Neerven and J.~Smith,
Phys.\ Rev.\ D {\bf 40},  54 (1989).

\bibitem{Altarelli:1979ub}
G.~Altarelli, R.~Ellis and G.~Martinelli,
Nucl.\ Phys.\ B {\bf 157}, 461 (1979). 

\bibitem{Altarelli:1977zs} 
G.~Altarelli and G.~Parisi,
Nucl.\ Phys.\ B {\bf 126}, 298 (1977).

\bibitem{Kang:2008wv} 
  Z.~B.~Kang, J.~W.~Qiu and W.~Vogelsang,
  Phys.\ Rev.\ D {\bf 79}, 054007 (2009)
  [arXiv:0811.3662 [hep-ph]].

\bibitem{Martin:2009iq} 
  A.~D.~Martin, W.~J.~Stirling, R.~S.~Thorne, G.~Watt,
  Eur.\ Phys.\ J.\ C {\bf 63}, 189 (2009)
  [arXiv:0901.0002 [hep-ph]].
  
\bibitem{Dittmaier:2014qza} 
  S.~Dittmaier, A.~Huss and C.~Schwinn,
  Nucl.\ Phys.\ B {\bf 885}, 318 (2014)
  [arXiv:1403.3216 [hep-ph]].  
  
\bibitem{Smith:1983aa} 
  J.~Smith, W.~L.~van Neerven and J.~A.~M.~Vermaseren,
  Phys.\ Rev.\ Lett.\  {\bf 50}, 1738 (1983).
  
\bibitem{Balazs:1995nz} 
  C.~Balazs, J.~w.~Qiu and C.~P.~Yuan,
  Phys.\ Lett.\ B {\bf 355}, 548 (1995)
  [hep-ph/9505203].  

\bibitem{D0:2013jba} 
T.~A.~Aaltonen {\it et al.}  [CDF Collaboration],
  Phys.\ Rev.\ D {\bf 89}, 072003 (2014)
  [arXiv:1311.0894 [hep-ex]];
  V.~M.~Abazov {\it et al.}  [D0 Collaboration],
  Phys.\ Rev.\ D {\bf 89}, 012005 (2014)
  [arXiv:1310.8628 [hep-ex]].  

\bibitem{Martin:2002aw} 
  A.~D.~Martin, R.~G.~Roberts, W.~J.~Stirling, R.~S.~Thorne,
  Eur.\ Phys.\ J.\ C {\bf 28}, 455 (2003)
  [hep-ph/0211080].

\bibitem{Bourrely:2001du} 
  C.~Bourrely, J.~Soffer and F.~Buccella,
  Eur.\ Phys.\ J.\ C {\bf 23}, 487 (2002)
  [hep-ph/0109160].

\bibitem{Bourrely:2013qfa}   
 C.~Bourrely, F.~Buccella and J.~Soffer,
  Phys.\ Lett.\ B {\bf 726}, 296 (2013)  [arXiv:1308.3567 [hep-ph]];
  C.~Bourrely and J.~Soffer,
  arXiv:1502.02517 [hep-ph].

\bibitem{Gluck:2000dy} 
  M.~Gl\"{u}ck, E.~Reya, M.~Stratmann and W.~Vogelsang,
  Phys.\ Rev.\ D {\bf 63}, 094005 (2001)
  [hep-ph/0011215];  
  Phys.\ Rev.\ D {\bf 53}, 4775 (1996)
  [hep-ph/9508347].

\bibitem{Gluck:2000ch} 
  M.~Gl\"{u}ck and E.~Reya,
  Mod.\ Phys.\ Lett.\ A {\bf 15}, 883 (2000)
  [hep-ph/0002182]; {\it see also:}  M.~Gl\"{u}ck, A.~Hartl and E.~Reya,
  Eur.\ Phys.\ J.\ C {\bf 19}, 77 (2001)
  [hep-ph/0011300].
  
\bibitem{deFlorian:2005mw} 
  D.~de Florian, G.~A.~Navarro and R.~Sassot,
  Phys.\ Rev.\ D {\bf 71}, 094018 (2005)
  [hep-ph/0504155].
  
\end{thebibliography}
\end{document}